\documentclass[aps,prx,twocolumn,amsfonts,showpacs,superscriptaddress]{revtex4} 
\usepackage{bm}
\usepackage{epsfig,amsopn}
\usepackage{graphicx,color}%%%%%%%%%%%%%%
\usepackage{amsmath,amssymb, amsfonts, amsthm}
\theoremstyle{definition}

\usepackage{braket}
\newcommand{\Tr}[0]{ \textrm{Tr}}
\newtheorem{theorem}{Theorem}
\newtheorem{lemma}{Lemma}

\newcommand{\eq}[1]{\begin{equation} #1 \end{equation}}
\newcommand{\eqa}[2]{\begin{equation} #1 \label{#2} \end{equation}}
\newcommand{\balign}[1]{\begin{align} #1 \end{align}}

\newcommand{\todayd}{\the\year/\the\month/\the\day}

\newcommand{\up}{\uparrow}
\newcommand{\down}{\downarrow}

\newcommand{\lb}{\label}
\newcommand{\nt}{\notag}

\newcommand{\eref}[1]{Eq.~\eqref{#1}}

\newcommand{\bel}{\begin{easylist}}
\newcommand{\eel}{\end{easylist}}
\newcommand{\be}[1]{\begin{enumerate} #1 \end{enumerate}}
\def \({\left(}
\def \){\right)}
\def \[{\left[}
\def \]{\right]}

\newcommand{\abs}[1]{\left|#1\right|}

\newcommand{\sumtwo}[2]%
{\mathop{\sum_{#1}}_{#2}}
\newcommand{\sumthree}[3]%
{\mathop{\mathop{\sum_{#1}}_{#2}}_{#3}}
\newcommand{\sumfour}[4]%
{\mathop{\mathop{\mathop{\sum_{#1}}_{#2}}_{#3}}_{#4}} 
\newcommand{\prodtwo}[2]%
{\mathop{\prod_{#1}}_{#2}}
\newcommand{\mintwo}[2]%
{\mathop{\min_{#1}}_{#2}}
\newcommand{\maxtwo}[2]%
{\mathop{\max_{#1}}_{#2}}
\newcommand{\maxthree}[3]%
{\mathop{\mathop{\max_{#1}}_{#2}}_{#3}}
\newcommand{\limtwo}[2]%
{\mathop{\lim_{#1}}_{#2}}
\newcommand{\suptwo}[2]%
{\mathop{\sup_{#1}}_{#2}}
\newcommand{\supthree}[3]%
{\mathop{\mathop{\sup_{#1}}_{#2}}_{#3}}
\newcommand{\supfour}[4]%
{\mathop{\mathop{\mathop{\sup_{#1}}_{#2}}_{#3}}_{#4}} 
\newcommand{\inftwo}[2]%
{\mathop{\inf_{#1}}_{#2}}
\newcommand{\infthree}[3]%
{\mathop{\mathop{\inf_{#1}}_{#2}}_{#3}}
\newcommand{\inffour}[4]%
{\mathop{\mathop{\mathop{\inf_{#1}}_{#2}}_{#3}}_{#4}} 
\newcommand\calA{{\cal A}}

\newcommand\calE{{\cal E}}
\newcommand\calF{{\cal F}}

\newcommand\calH{{\cal H}}
\newcommand\calI{{\cal I}}
\newcommand\calJ{{\cal J}}
\newcommand\calK{{\cal K}}

\newcommand\calT{{\cal T}}

\newcommand{\ep}{\varepsilon}

\def\QED{\mbox{\rule[0pt]{1.5ex}{1.5ex}}}

\def\endproof{\hspace*{\fill}~\QED\par\endtrivlist\unskip}

\newenvironment{proofof}[1]{\vspace*{5mm} \par \noindent
{\bf Proof of #1:\hspace{2mm}}}{\endproof
}

\begin{document}
\title{Coherence cost for violating conservation laws}
\author{Hiroyasu Tajima}
\affiliation{Yukawa institute for theoretical physics, Kyoto University
Oibuncho Kitashirakawa Sakyo-ku Kyoto city, Kyoto, 606-8502, Japan }

\author{Naoto Shiraishi}
\affiliation{Department of Physics, Gakushuin University, 1-5-1 Mejiro, Toshima-ku, Tokyo, 171-8588, Japan} 

\author{Keiji Saito}
\affiliation{Department of Physics, Keio University, 3-14-1 Hiyoshi, Yokohama, 223-8522, Japan}

\begin{abstract}
Nature imposes many restrictions on the operations that we perform. Many of these restrictions can be interpreted in terms of {\it resource} required to realize the operations. Classifying required resource for different types of operations and determining the amount of resource are the crucial subjects in physics. Among many types of operations, a unitary operation is one of the most fundamental operation that has been studied for long time in terms of the resource implicitly and explicitly. Yet, it is a long standing open problem to identify the resource and to clarify the necessary and sufficient amount of resource for implementing a general unitary operation under conservation laws. In this paper, we provide a solution to this open problem. We derive an asymptotically exact equality that clarifies the necessary and sufficient amount of quantum coherence as a resource to implement arbitrary unitary operation within a desired error. In this equality, the required coherence cost is asymptotically expressed with the implementation error and the degree of violation of conservation law in the desired unitary operation. We also discuss the underlying physics in several physical situations from the viewpoint of coherence cost based on the equality. This work does not only provide a solution to a long-standing problem on the unitary control, but also clarifies the key question of the resource theory of the quantum channels in the region of resource theory of asymmetry, for the case of unitary channels. 
\end{abstract}

\pacs{
03.65.Ta,	%Foundations of quantum mechanics; measurement theory
03.67.-a, %Quantum information
05.30.-d, %Quantum statistical mechanics
42.50.Dv, %Quantum state engineering and measurements
}

\maketitle
\section{Introduction}
The laws of physics impose many limitations on the operations that we perform. For example, the thermodynamic second law imposes a restriction on the amount of work done by heat engine. Quantum speed limit \cite{speed-limit}, which originates from the time-energy uncertainty relation imposes a fundamental limitation on the speed of quantum operation \cite{Lloyd2000}. The Wigner-Araki-Yanase theorem, an important theorem on quantum measurement, imposes the restriction that instantaneous values of time-varying physical quantities cannot be recorded without errors \cite{Wigner1952,Araki-Yanase1960,Yanase1961,OzawaWAY}.

Most of such limitations can be interpreted in terms of the {\it resource} required to realize the desired operation. In the examples above, the quantum speed limit can be understood that if we wish to change a quantum state quickly, we must prepare large energy fluctuations as a resource. The Wigner-Araki-Yanase theorem can be understood as an assertion that if we wish to accurately measure time-varying physical quantities under the energy conservation law, we must prepare considerable energy fluctuations as a resource \cite{OzawaWAY}.

Recently, the resource theory in the quantum channels has cast new light on many types of quantum operations in terms of resources. In the resource theory, some operations are classified as free operations, and some states which cannot be obtained through free operations are classified as resource states. The task in the resource theory of quantum channels is to realize a desired operation which is not free, by combining free operations and resource states. Estimating the amount of resource required to implement the desired operation is a central subject in this area. So far, this subject has been studied for various classes of the free operations, such as quantum thermodynamics \cite{Renner-channels, Brandao-channels}, resource erasure \cite{Winter-channels} and incoherent operations \cite{Winter-channels0}.

In this paper, we address a general unitary operation, and unveil the underlying physics in the viewpoint of resources. Since a unitary operation is one of the most fundamental operations in physics, there are many relevant researches \cite{jaynes1963,optics1,optics2,optics3,optics4,optics5,optics6,optics7,catalyst, Woods, Malabarba, ozawa1, ozawa2,Karasawa2007,Karasawa2009,TSS}. For instance, based on specific models in the quantum optics such as the Jaynnes-Cummings model, sufficient conditions on a certain resource for a desired unitary dynamics have been analyzed \cite{jaynes1963,optics1,optics2,optics3,optics4,optics5,optics6,optics7}. These results were further deepened by \r{A}berg in a more general framework and a sufficient conditions for an arbitrary unitary dynamics are argued \cite{catalyst}. This framework is applied to various objects such as clock \cite{Woods} and heat engine \cite{Malabarba}.

The necessary condition to implement the desired unitary dynamics have been also studied intensively \cite{ozawa1,ozawa2,Karasawa2007,Karasawa2009,TSS}. The first relevant study goes back to Ozawa's work in 2002, which addressed the limitations on the quantum computation due to the conservation laws \cite{ozawa1}. Ozawa considered a unitary gate on the target system via the spin-preserving interaction between the target system and an external quantum system that corresponds to an external apparatus. Under this setting, he used the Wigner-Araki-Yanase theorem and demonstrated that implementation of Controlled-NOT gate within a small error requires large amount of fluctuation in the external apparatus as a resource \cite{ozawa1}. The generalization of this research has been actively studied \cite{ozawa2,Karasawa2007,Karasawa2009,TSS}. In particular, in Ref.\cite{TSS}, it was shown that the fluctuation in the external apparatus must be a {\it quantum fluctuation} (i.e., quantum coherence). Hence, the required resource for unitary operations under the conservation law can be identified as the {\it quantum coherence} of the conserved quantity.

\begin{figure}
\includegraphics[width=9cm]{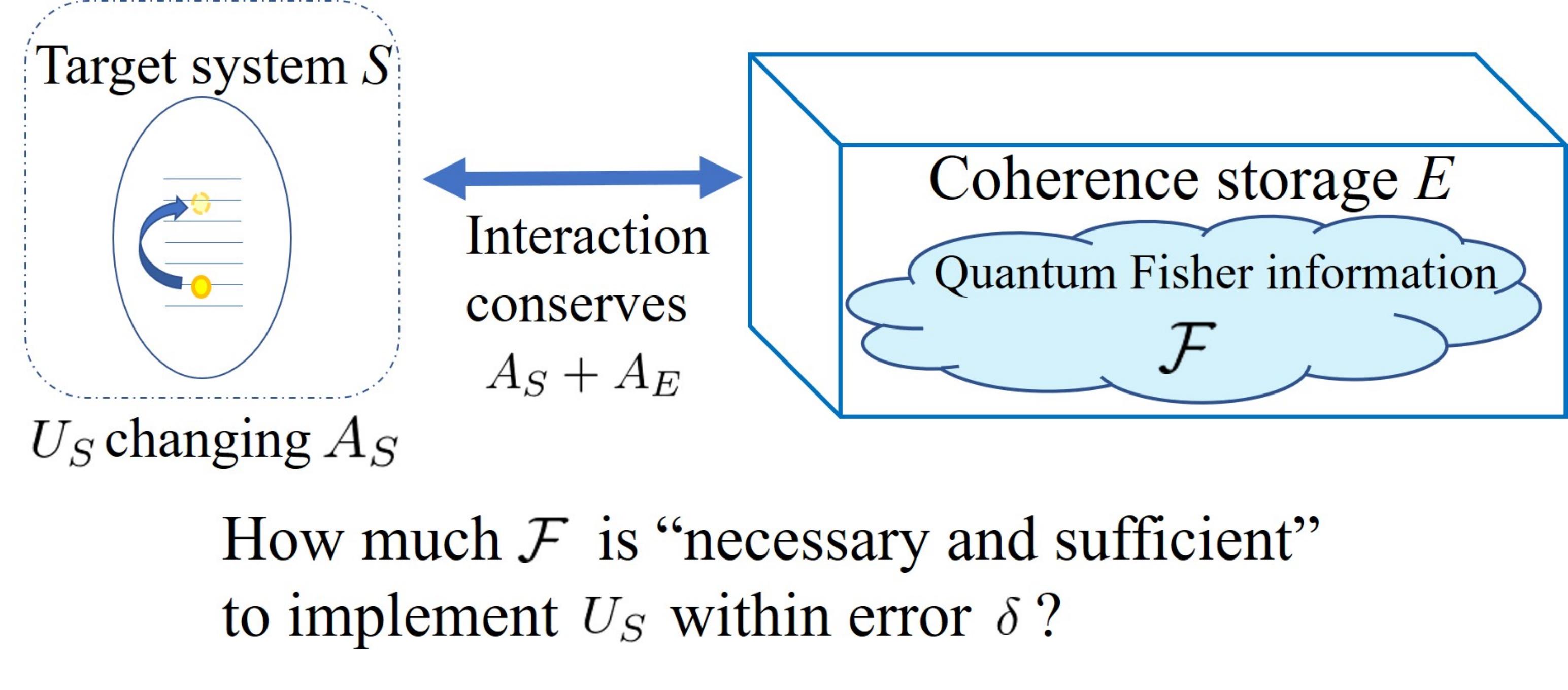}
\caption{Schematic of the unitary operation by attaching the system (S) to the external system (E). Conserved quantities exist in the system, of which one of them is denoted by $A_S$. The external system is supposed to exhibit the same type of quantity $A_E$, and the operator $A_S + A_E$ is a conserved quantity for the time-evolution of the entire composite system. We aim to implement a unitary operation into the system such that the desired unitary operator does not conserve the quantity $A_S$. Hence, the quantity $A_E$ in the external system must compensate the change of $A_S$. In general, to create the unitary dynamics for the system, it is natural to use quantum coherence inside the external system. Our purpose is to estimate the quantum coherence to implement the desired quantum coherence.}
\label{setup}
\end{figure}

In spite of these progress, we do not still reach an exact solution on the necessary and sufficient amount of quantum coherence for a general unitary operation, which is the most critical goal in this subject. Herein, we present a complete solution on this long-standing problem. Following the same framework as in the previous researches, we regard the unitary operation in the target system as a physical phenomenon that results from the interaction between the target system and the external system (external apparatus). See the schematic in Fig.\ref{setup}. Then we consider the amount of quantum coherence in the external system for realizing the unitary time evolution in the target system. We show a simple and asymptotically exact equality that clarifies the necessary and sufficient amount of quantum coherence to implement arbitrary unitary dynamics within the desired error. This asymptotic equation links three fundamental quantities: the implementation error, amount of coherence, and degree of asymmetry (violation of the conservation law) of the desired unitary operation. These findings do not only give a solution to a long-standing problem but also clarify the key question of the resource theory of the quantum channels in the region of the resource theory of asymmetry \cite{Bartlett, Gour, Marvian, Marvian2016, Marvian-thesis,Takagi2018, Marvian2018, Lostaglio2018}.

This paper is organized as follows. In Sec. $2$, we explain the setup and define the amount coherence, the error of the desired operation onto the system, and the asymmetry that is a measure of degree of violation of conservation law. In Sec. $2$, we show the main results on the relations between several quantities defined in Sec. $1$. In Sec. $4$, we discuss the underlying physics behind several quantum manipulations. 
In Sec. $5$, we extend our result to $G$-covariant operation, and show that our results clarifies the amount of necessary and sufficient resource to implement non-free unitary, in the case of resource theory of $U(1)$-asymmetry.
In Sec. $6$ and $7$, we briefly present the proof of the main results in Sec. $2$. Finally, conclusions are drawn in Sec. $8$.

\section{Setup and aim}
We consider a system which is symbolically denoted by $S$. We assume that the system contains a finite dimension of the Hilbert space ${\cal H}_S$. 
Suppose that the system comprises a conserved quantity, which commutes with the system's Hamiltonian. 
In the above supposition, we
%KS include
also count
the system's Hamiltonian as a conserved quantity,
%KS
since an energy is conserved.
%, implying conservation of energy. 
We take one of conserved quantities and denote it by $A_S$.

We consider a mechanism to implement the desired unitary operation $U_S$ that does not commute with the conserved quantity $A_S$.
Obviously, the usual time-evolution driven by the static system's Hamiltonian cannot generate such a unitary operation. Hence, we resort to an external coherence resource to create such a unitary operation.
To this end, we attach the external system to the system and use the coherent dynamics of the composite system. See the schematics in Fig.\ref{setup}. 
%It is nontrivial whether the system's reduced density matrix obeys unitary dynamics. 
We may have in mind that in experimental situations, an experimental apparatus that induces the time-dependent Hamiltonian of the system plays the role of the external system. 
In this study, we consider the fundamental limitation to implement the desired unitary operation on the target system in the composite-system setup. 
From the viewpoint of quantum information theory, we address the following problem: what is the fundamental limitation to implement the unitary dynamics violating the conservation law using the coherence resource stored in the external system ? 
In particular, we aim to clarify the necessary and the sufficient amount of coherence resource to implement a unitary operation.
%In more detail, we aim to clarify the relation between the amount of coherence in the external system and implementation error of the unitary operation, and find a certain between possible and impossible unitary operation for a given quantum coherence in the external system.

We denote the external system by $E$, and the Hilbert space of the external system by ${\cal H}_E$.
The external system is attached to the target system through an interaction Hamiltonian, and the total composite system $S+E$ is assumed to be isolated. Let $H_{SE}$ be a Hamiltonian of the total composite system, and consider that the entire composite system evolves in time with this total Hamiltonian. The most crucial assumption in our setup is that the same class of physical quantity as $A_S$ exists in the external system which is denoted by $A_E$, and the sum of operators $A_S + A_E$ is conserved ~\cite{footnote-1}, i.e.,
\begin{align}
\left[ H_{SE},A_S+A_E \right] &=0 \, . \label{condasae}
\end{align}
For the case that we take energy as a conserved quantity, i,e., $A_S$ and $A_E$ are the Hamiltonian of the target system and external system, this assumption implies that the interaction between the target system and the external system does not store energy.

We consider the case that the initial state of the whole system is a product state of the target system and the external system.
We now introduce the time evolution operator $\Lambda_S$ of the target system for a finite time interval $\tau$:
\begin{align}
\Lambda_{S}(\rho_S):=\Tr_{E}\left[ e^{-i \tau H_{SE} /\hbar} (\rho_S\otimes\rho_E) e^{i \tau H_{SE}/\hbar} \right] \, , \label{ls}
\end{align}
%\vspace{5pt}
where $\rho_S$ and $\rho_E$ are the initial states of the target system and the external system, respectively. 
This operator $\Lambda_S$ is a completely positive and trace preserving (CPTP) map that maps a reduced density matrix of the system from the initial time to that in the time $\tau$. 
%It is crucial to recognize that for a given system's initial state, the time evolution of the target system is determined by the initial density matrix spanned in the external system and the unitary time-evolution for the total composite system. 
%Note here that if the CPTP map is a desired unitary map, the operation is independent of the initial state of the system $\rho_S$. 
We herein aim to implement a unitary operation that violates the conservation law in the target system. 
In our setup, the quantity $A_S+A_E$ is conserved from the condition (\ref{condasae}); hence, in the dynamics the quantity $A_E$ must compensate the change in $A_S$.

The implementation of $U_S$ is fully determined by the following set
\begin{align}
{\cal I} & :=({\cal H}_E, A_E, \rho_E, e^{-i \tau H_{SE}/\hbar}) ~\, ,
\end{align}
which we call the {\it implementation set} for the unitary operation. When $\Lambda_S (\rho_S)$ approximates $U_S \rho_S U_S^{\dagger}$ accurately for an arbitrary initial density matrix $\rho_S$, we say that the set ${\cal I}$ is a {\it good} implementation set.

To quantify the accuracy of implementation, we introduce a measure on the error in the implementation of the desired unitary dynamics. To this end, we use the entanglement Bures distance~\cite{hayashi} defined as 
\begin{align}
L_{e}(\rho_S,\Lambda ) &:=\sqrt{2(1-F_{e}(\rho_S,\Lambda))},\label{uerror}\\
F_{e}(\rho_S,\Lambda) &:=\sqrt{\bra{\psi}_{SR} [1_{R}\otimes\Lambda](\psi_{SR}) \ket{\psi}_{SR}},
\end{align}
where $\ket{\psi}_{SR}$ is a purification of $\rho_{S}$ and $R$ stands for the reference space. The operation $[1_{R}\otimes\Lambda](\psi_{SR})$ is an abbreviation of $[1_{R}\otimes\Lambda](|\psi_{SR}\rangle \langle \psi_{SR}|)$. Throughout this paper, we frequently use this abbreviation. The operator $\Lambda$ is an arbitrary time-evolution operator that acts only on the Hilbert space of the target system. The fidelity $F_{e}(\rho_S,\Lambda)$ provides an amplitude of the overlap between the initial state and the final state driven by the time-evolution operator $\Lambda$. The Bures distance $L_{e}(\rho_S,\Lambda)$ quantifies the distance between these two states. 
We set the time-evolution as $\Lambda =\Lambda_{U^{\dagger}_S}\circ\Lambda_S$, where $\Lambda_S$ is the CPTP map defined in Eq.(\ref{ls}) and $\Lambda_{U^{\dagger}_S}$ is an inverse time-evolution of the desired unitary dynamics, i.e., 
\eq{
\Lambda_{U^{\dagger}_{S}}(\rho_S):=U^{\dagger}_{S}\rho_S U_{S} \, .
}
The operator $\Lambda$ is a successive application of these two maps.
Subsequently, the Bures distance measures the distance between the final state driven by the desired unitary time-evolution and the actual final state, which serves as a measure of error.
We write this error as a function of the initial state $\rho_S$ as
\eq{
\delta_{}(\rho_S):=L_{e}(\rho_S,\Lambda_{U^{\dagger}_S}\circ\Lambda_S ) \, ,\label{def-error-state}
}
and we define the accuracy of the implementation as the worst case within all initial states: 
\begin{align}
\delta_{\calI}:=\max_{\rho_S}\delta(\rho_S).\label{implementation-error}
\end{align}
If the error $\delta_{\calI}$ of an implementation set ${\cal I}$ is less than a value $\delta$, we say that {\it the implementation set ${\cal I}$ realizes $U_S$ within error $\delta$} and express it as $\calI\models_{\delta} U_S$.

Next we introduce the measure of the amount of coherence in the external system $E$.
We use the quantum Fisher information with respect to $A_E$ \cite{Fisher-coherence1,Q-Fisher,Q-Fisher=Q-fluctuation1,Q-Fisher=Q-fluctuation2, Marvian-nogo} defined for a given state $\rho$ as
\begin{align}
\calF(\rho) &:=\min_{\{q_j,\phi_j\}: \rho=\sum_{j}q_j\phi_j} 4 \sum_jq_j V^2_{A_E }(\phi_j )\, . \label{F=Q}
\end{align}
Here the minimum value is searched over all possible decompositions $\{q_j,|\phi_{j}\rangle \}$ of the fixed density matrix $\rho$, and $V_{A_E}(\rho)$ is the standard deviation of the quantity $A_E$ for the pure state $| \phi_j\rangle $, i.e., 
\eq{
V_{A_E}(\phi_j ) :=\sqrt{\langle \phi_j | A_E^2 | \phi_j \rangle -\langle \phi_j | A_E |\phi_j \rangle^{2} } \, .
}
If the decomposition $\phi_{j}$ for the density matrix is identical to the eigenstates of $A_E$, the Fisher information $\calF$ is exactly zero. We regard this case as {\it classical}. 
The Fisher information $\calF$ takes a finite value when $|\phi_{j} \rangle$ is a superposition of eigenstates with different eigenvalues of $A_E$. In particular, if $\rho$ is a pure state given by $|\phi \rangle \langle \phi |$ , the quantum Fisher information is equal to $4V_{A_E}(\phi )$. Therefore, the quantum Fisher information is interpreted as a measure of the quantum fluctuation of $A_E$. We note that the quantum Fisher information can be also expressed as
\begin{align}
\calF(\rho)=2\sum_{a,b}\frac{(p_a-p_b)^2}{p_a+p_b}|A_{ab}|^2 \, , \label{afi}
\end{align}
%KS
where $p_a$ is the $a$-th eigenvalue of the density matrix $\rho$ with the eigenvector $\psi_a$, and $A_{ab}=\bra{\psi_a}A_E\ket{\psi_b}$. 

Regarding the quantum Fisher information as an amount of coherence that the external system contains, we consider the amount of coherence cost that the external system must bear to materialize the desired unitary operation into the system. We consider the situation where the desired unitary operation is achieved within the error $\delta$. We define the coherence cost $\calF_{U_S, \delta}$ as the minimal value of the quantum coherence within all possible implementation sets that implements the desired unitary operation within error $\delta$:
\begin{align}
\calF_{U_S, \delta}:=
\min_{\calI\models_{\delta}U_S}\calF(\rho_E). \label{deff}
\end{align}

We finally define the degree of asymmetry. The asymmetry in the present context implies a degree of violating the conservation law inside the target system by the unitary operation $U_S$. We quantify this through the amount of noncommutativity between $U_S$ and $A_S$:
\begin{align}
\calA_{U_S}:=\frac{\lambda_{\max}(A'_S-A_S)-\lambda_{\min}(A'_S-A_S)}{2}, \label{asym}
\end{align}
where $A'_S :=U^{\dagger}_SA_SU_S$, and $\lambda_{\max}(X)$ and $\lambda_{\min}(X)$ are the maximum and minimum eigenvalues of the operator $X$, respectively.
By construction, $\calA_{U_S}$ is non-negative, and becomes $0$ if and only if $U_S$ and $A_S$ commute with each other \cite{footnote0}.
Hence it is clear that a finite value of the asymmetry reflects the violation of the conservation law by the unitary operation. 

Our framework and results can be easily extended to the framework of resource theory of asymmetry \cite{Bartlett, Gour, Marvian, Marvian2016, Marvian-thesis,Takagi2018, Marvian2018, Lostaglio2018} that is a fruitful region in the resource theory.
In the extension, our results clarify necessary and sufficient amount of resource to implement arbitrary unitary operations under $G$-covariant operations. It is an answer to the key question of resource theory of quantum channels, in case of the resource theory of asymmetry.
We discuss this extension in Sec. \ref{sectionGcov}.

%We note that the quantity $\calA_{U_S}$ can be rewritten as
%\begin{align}
% \calA_{U_S}&=\max_{\rho_S} \sqrt{ {\rm tr} ( ( A'_S-A_S )^2 \rho_S ) - \left[ {\rm tr} ( (A'_S-A_S) \rho_S) \right]^2 } \, , \label{w-fluc} \\
% A'_S & :=U^{\dagger}_SA_SU_S.
%\end{align}
%This expression means that $\calA_{U_S}$ quantifies the fluctuation of the change in $A_S$ through the unitary operation $U_S$.
%The quantity $\calA_{U_S}$ has a close connection with the maximum expected value of the change in $A_S$:\begin{align}\calA_{U_S}\le \max_{\rho_S}|\Tr[\rho_S(A'_S-A_S)]| \le2\calA_{U_S}.\label{w-max} \end{align}

\section{Main Results}
\subsection{Coherence cost for unitary operations}

We present the main results of this study and discuss their crucial physical consequences. The proof of the results will be provided later. 
%When we specify the regime of error of the unitary operation, one can obtain the relations described by the following two theorems.
\begin{theorem}\label{T1}
Let $\delta$ be a real positive value satisfying $0\le\delta\le\sqrt{2}$.
For any implementation set $\calI$ satisfying $\delta_{\calI}\le\delta$, the following inequality holds
\begin{align}
\sqrt{\calF(\rho_E) }\ge\frac{\calA_{U_S}}{\delta }-4\|A_S\|.\label{T1-2}
\end{align}
\end{theorem}
\begin{theorem}\label{T2}
Let $\delta$ be a real positive value satisfying $0\le\delta\le4\sqrt{2}\calA_{U_S}/(9\|A_S\|)$. For an arbitrary value $\calF$ satisfying the following inequality, there exists an implementation set $\calI$ that satisfies $\delta_{\calI}\le\delta$ and $\calF(\rho_E)=\calF$
\begin{align}
\sqrt{\calF}\ge\frac{\calA_{U_S}}{\delta}+\sqrt{2}\|A_S\|.\label{T2-1}
\end{align}
\end{theorem}
Theorem \ref{T1} is the inequality on the necessary condition required for {\it any} implementation set, while Theorem \ref{T2} guarantees the {\it existence} of at least one implementation set that realizes the desired unitary operation, provided that the inequality is satisfied. In section \ref{sec:theot2}, we present the existence of such an implementation set constructively. 
We emphasize that we do not impose any conditions on the unitary operation and hence the two theorems hold for any desired unitary operation.

These two theorems bound the amount of coherence cost both from below and above:
Thus, as shown below, they provide information on the necessary and sufficient amount of coherence in the external system within a given operational error. 
In particular, if the operation error asymptotically vanishes, the necessary amount and the sufficient amount of coherence become asymptotically equal. 
%To derive this, we consider the two theorems in more depth by considering the product of the operation error and the quantum Fisher information. 
To clarify the fact that Theorem \ref{T1} is a necessary condition, we take its contraposition.
The contraposition states that it is \textit{impossible} to find a implementation set $\calI$ satisfying $\delta_{I}\le\delta$ and $\calF(\rho_E)=\calF$ if the tuple of positive numbers ($\calF$,$\delta$) obeys the following inequality
\begin{align}
\sqrt{\calF}<\frac{\calA_{U_S}}{\delta}-4\|A_S\|. \label{A}
\end{align}
Meanwhile, Theorem \ref{T2} is the sufficient condition.
It guarantees that it is \textit{always possible} to obtain at least one implementation set $\calI$ satisfying $\delta_{I}\le\delta$ and $\calF(\rho_E)=\calF$ if the tuple ($\calF$,$\delta$) obeys the following inequality
\begin{align}
\sqrt{\calF}\ge\frac{\calA_{U_S}}{\delta}+\sqrt{2}\|A_S\| \label{B}
\end{align}
for the region of $0\le\delta\le4\sqrt{2}\calA_{U_S}/(9\|A_S\|)$.

\begin{figure}
\includegraphics[width=7.5cm]{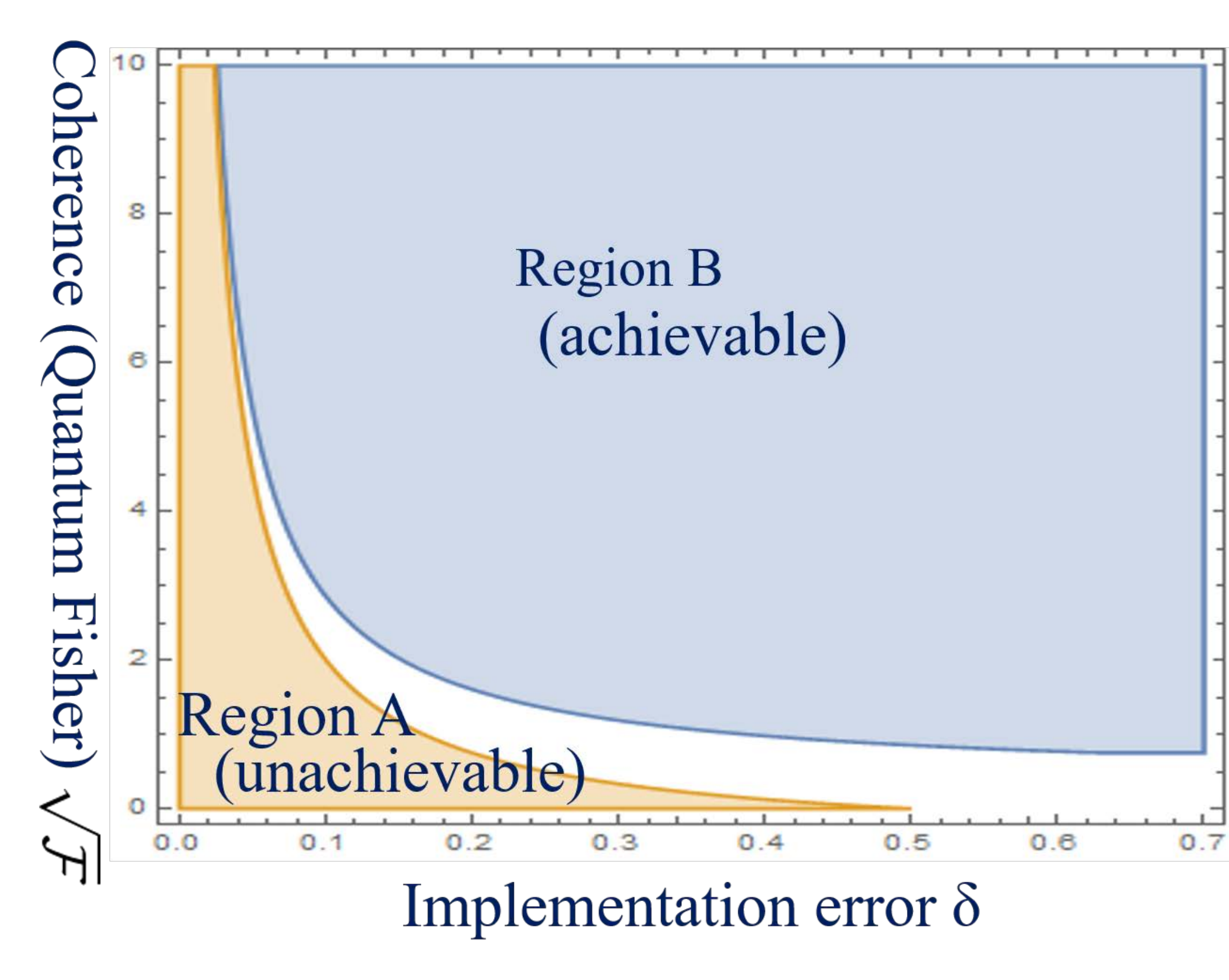}
\caption{
A graph indicating \eqref{A} and \eqref{B} for the specific model.
The system is a qubit system whose Hamiltonian is $(1/2)(\ket{1}\bra{1}-\ket{0}\bra{0}$ and the desired unitary operation is the bit-flip unitary $\ket{1}\bra{0}+\ket{0}\bra{1}$. 
In region $A$, no sets exist to achieve the unitary operation indicated from the inequality \eqref{A}, while in region $B$ at least one set exists to achieve the unitary operation from the inequality \eqref{B}.} 
\label{Graph}
\end{figure}

We numerically demonstrate the obtained bounds by taking a specific example of the qubit system whose Hamiltonian is $(1/2)(\ket{1}\bra{1}-\ket{0}\bra{0})$ and the desired unitary operation is the bit-flip, i.e., $U_S=\ket{1}\bra{0}+\ket{0}\bra{1}$. 
If the inequality \eqref{A} is satisfied, one can obtain the regime of the operation error and the quantum Fisher information that cannot achieve the desired unitary operation within a given error using the quantum Fisher information. 
We depict the unachievable regime indicated by the region $A$ in Fig.~\ref{Graph}. In region $A$, one cannot implement the desired unitary operation within error $\delta$ by \textit{any} implementation set $\calI$ satisfying $\calF(\rho_E)=\calF$. 
That is, in the region $A$, the amount of coherence $\calF$ is insufficient to implement the desired unitary operation within the error $\delta$. 
Meanwhile, if the inequality \eqref{B} is satisfied, we have at least one implementation set to achieve the unitary operation within a given error, whose regime is shown as the region $B$ in Fig.~\ref{Graph}. (One might wonder why the region $B$ extends to the region $\delta>4\sqrt{2}\calA_{U_S}/(9\|A_S\|)$.
Note that even in the region of $\delta>4\sqrt{2}\calA_{U_S}/(9\|A_S\|)$, Theorem \ref{T2} guarantees that it is possible to find at least one implementation set $\calI$ satisfying $\delta_{\calI}\le\delta$ and $\calF(\rho_E)=(17\|A_S\|)/(4\sqrt{2}\|A_S\|)$, by substituting $\delta=4\sqrt{2}\calA_{U_S}/(9\|A_S\|)$ in \eqref{B}.)
%HT rewrite the following three sentences
We also depict the achievable regime indicated by the region $B$.
In the region $B$, we can implement the desired unitary operation within the error $\delta$ using an implementation set $\calI$ satisfying $\calF(\rho_E)=\calF$. That is, in the region $B$, the amount of coherence $\calF$ is sufficient to implement the desired unitary within the error $\delta$.

As illustrated by the specific model shown in the figure, the two regimes (achievable and unachievable regimes) converge to the same line as the operation error goes to zero. This implies that the minimum coherence cost $\calF_{U_S,\delta}$ approaches the bound in Theorem \ref{T1}.
The relations (\ref{A}) and (\ref{B}) lead to that this behavior is general, and hence we eventually arrive at the following asymptotic relation for the coherence cost.
\begin{align}
\sqrt{{\cal F}_{U_S,\delta}} =\frac{\calA_{U_S}}{\delta}+O(\|A_S\|) ~~{\rm for}~~\delta\to 0 \, .\label{maineq}
\end{align}
The two inequalities (\ref{T1-2}) and (\ref{T2-1}) as well as the asymptotic equality (\ref{maineq}) explicitly show a close relation among the coherence cost, degree of asymmetry, and error of unitary operation. 
For a fixed asymmetry, a large coherence is necessary for an accurate unitary operation. The equation represents the trade-off relation between them. 
Also, a large asymmetry requires a large coherence. 
%These relations explicitly indicate that the target system really {\it requires} a coherence of the external system for the unitary operation.

The asymptotic equality (\ref{maineq}) shows that the coherence cost linearly depends on the asymmetry.
From the definition (\ref{asym}), the asymmetry $\calA$ is of the same order of magnitude as the conserved quantity $A_S$. 
Suppose that the system of interest is a macroscopic system, and the conserved quantities satisfy the extensivity with respect to the system size.
In this case, the coherence cost for achieving the desired operation must also satisfy the extensivity with respect to the system size. 
This is suggestive, as it appears to imply difficulties in the implementation of unitary operations for macroscopic systems because a significant coherence cost is required.

\subsection{Coherence cost for a fixed initial state}

In the previous subsection, we established an asymptotically tight relation among the coherence cost, error of unitary operation and amount of asymmetry. Note that the unitary operation is applied for any initial state in the system. 
Meanwhile, suppose that we are interested in a transformation from a certain fixed initial state to a desired final state. 
%Subsequently, we only need to consider the transformation between given fixed states. 
In this case, what conditions must be imposed on the coherence cost in the external systems ? We herein address this question. 
Let $\{\psi_i\}$ be an orthonormal basis of the system, and let $U_S$ be a transformation providing the desired final state from the initial state $\rho_S$. 
We herein discuss the properties of transformation on the implementation set $\calI$. We have the following inequality similar to \eref{T1-2}:
\begin{align}
\sqrt{\calF(\rho_E)}\ge\frac{\chi(\rho_S, \{ \psi_i\} )}{5\sqrt{\delta_{}(\rho_S)^2+\sum_ir_i\delta_{}(\psi_i)^2}}-4\|A_S\|,\label{true}
\end{align}
where $r_{i}:=\braket{\psi_i|\rho_S|\psi_i}$ quantifies the weight of $\ket{\psi_i}$ in $\rho_S$.
We also defined the fluctuation of the change in $A_S$ with respect to the basis $\{ \psi_i\}$:
\eq{ \chi(\rho_S, \{ \psi_i\} ):=\sqrt{\sum_{i} r_{i}(\left<A'_S-A_S\right>_{\psi_i}-\left<A'_S-A_S\right>_{\rho_S})^2 } \, , 
} 
where $\langle...\rangle_{\psi_i}={\rm tr}(...|\psi_i \rangle \langle \psi_i | )$ and $\langle...\rangle_{\rho_S}={\rm tr}(...\rho_S )$.
The inequality \eqref{true} implies that a large coherence is required even when the unitary transformation is applied to a limited initial state.

We emphasize that the coherence cost implied by the inequality is valid in implementing many {\it non-unitary} CPTP maps. 
An example includes a gate on a $d$-level system that behaves as a bit-flip unitary only for the space spanned by the ground state and the first excited state but behaves as a non-unitary gate for other states. 
Because this is not a unitary operation for the entire Hilbert space, we cannot apply Theorem \ref{T1}; however, the inequality \eqref{true} is available to estimate the required coherence cost. 
This implies that to implement the gate with a small error, the required coherence must be inversely proportional to the error. 

\section{Application}
In this section, based on the relations (\ref{maineq}) and (\ref{true}), we discuss the underlying physical mechanisms in manipulating a quantum state.
For several cases, a special attention is paid on the coherence cost that must be prepared in the external system.
The cases in the subsections A and B are discussed in terms of the relation (\ref{maineq}) and the case in the subsection C is
discussed with the relation (\ref{true}).

\subsection{Underlying physics to implement time-dependent Hamiltonian}
Our results connect the coherence cost to asymmetry in implementing a unitary operation. A typical conserved quantity in nature is energy; hence, our theory is most importantly applicable for implementing a time-dependent Hamiltonian that changes the energy inside the system, such that the following unitary operation is applied
\begin{align}
U_S &:=\calT \exp \left( -i\int_0^\tau dt\, \tilde{H}_S(t) \right) \, , \label{us}
\end{align}
where $ \tilde{H}_S(t)$ is a time-dependent Hamiltonian, and $\calT$ implies the time-ordered product.

Implementing the time-dependent Hamiltonian is crucial in many experiments to manipulate of quantum states, such as qubit manipulation. In such experiments, the external system is merely an experimental apparatus. Experimental apparatus typically use the classical electromagnetic interaction with the target system. From the asymptotic equality (\ref{maineq}), a perfect unitary operation implies that the external system provides quantum coherence. Indeed, classical electromagnetic fields are obtained in the limit of large amplitude of the coherent state. Hence it is consistent with the present theory.

%KS It is noteworthy
Note
that the quantum Fisher information is connected to the variance as $\langle (A-\langle A \rangle_{\rho})^2 \rangle_{\rho} \ge {\cal F}(\rho)/4$, where
$\langle ...\rangle_{\rho}={\rm tr}(...\rho )$. Combining this with the asymptotic equality (\ref{maineq}), the perfect unitary operation implies the divergence of variance in the external system. This can be interpreted as follows. When the desired unitary control changes the quantity $A_S$ in the system, the amount of change must be compensated by $A_E$. Meanwhile, to complete the operation, the state in the external system should not be damaged by the change in $A_E$. This is possible if the external state has a large variance with respect to the the physical quantity $A_E$ \cite{footnote3}, as a small change in the quantity $A_E$ can be negligible compared with a large fluctuation. 
Hence, a large variance is advantageous for generating a unitary operation for the target system.

\subsection{Quantum heat engines: quantum work storage}\label{ApplicationB}

In particular, the application of our results to quantum cyclic heat engines is suggestive.
In the analysis of quantum heat engines, a model called the {\it standard model} is widely used \cite{Ehrenfest,A-Lenard, Kurchan,Tasaki,seifert-rev,Park2013,ikeda2015,review,Shiraishi,Park2017}. Here, we set the composite system of the working body and heat bath as the target system and consider the role of the external system. 
To this end, we consider an expectation value of quantum work that is extracted from the target system. This is defined by the difference of the energy in the target system:
\begin{align}
\left<W\right>:=\Tr[\rho_S\tilde{H}_{S}(\lambda(0))-U_S\rho_SU^{\dagger}_S\tilde{H}_{S}(\lambda(\tau))]\label{standardwork},
\end{align}
where $\lambda (t)$ is the control parameter of the system Hamiltonian. In the standard model, the work extracted from the target system is considered to be stored in an external work storage through the back action of control parameters. 
The work storage is
% KS merely
nothing but
the external system in the present language. 
In a typical experimental setup, the work storage (external system) is an experimental apparatus that controls the parameters \cite{footnote4}. In several theoretical setups \cite{Malabarba,Popescu2014,Popescu2015,HO,brandao2015,
SSP,TWO,oneshot3,gourreview,MTH,Tasaki2015}, the work storage is prepared to store the work done. 
According to Theorem \ref{T1}, the external system must have sufficiently large energy fluctuations when realizing a time-dependent Hamiltonian. As argued in the previous subsection, for a perfect unitary control on the system, the required energy fluctuation must be large such that the energy gain of the work storage is negligibly small compared to the energy fluctuation of the work storage.

The above discussion provides a suggestive message regarding the {\it detectability} of work of quantum heat engines, which is analyzed in the previous results \cite{m-based,Marti}.
Again, we consider the heat engine that is a composite system of working body and heat baths, and an external work storage that stores the work extracted from the heat engine.
Let us consider a situation in which we wish to detect the energy gain in the work storage by measuring the work storage.
This situation corresponds, for example, to determining the amount of work by comparing the initial and final positions of the weight lifted by the heat engine.
Using the trade-off relation between information gain and disturbance in measurements, the previous studies \cite{m-based,Marti} have shown that if the time evolution of the heat engine can be described in terms of unitary dynamics as assumed by the standard model, the amount of work cannot be detected by measuring the storage. 
Our result provides an intuitive explanation to why such a loss of detectability occurs.
As discussed above, the work storage must exhibit a much greater energy fluctuation than the energy gain from the heat engine.
Therefore, the energy gain of the work storage is swallowed by its energy fluctuation; thus, we cannot determine the amount of energy gain.

\subsection{Coherence cost for entanglement erase}\label{entanglement-erase}

Our main result \eqref{maineq} provides the coherence cost for the implementation of unitary gates.
However, we can evaluate the coherence costs for the state transformations other than unitary.
As a typical example, we apply our results to entanglement erasure.
Given an entangled initial state in the form $\alpha\ket{00}+\beta\ket{11}$ with arbitrary $\alpha, \beta$ satisfying $\abs{\alpha}^2+\abs{\beta}^2=1$, we perform the following entanglement erase process:
\begin{align}
\alpha\ket{00}+\beta\ket{11}\rightarrow\alpha\ket{00}+\beta\ket{10}\label{e-erase}
\end{align}
This erasure process might be a non-unitary CPTP operation because the state transformation for initial states other than the given form is not specified.
Therefore, the formula \eqref{maineq} cannot evaluate the necessary amount of the coherence for the state transition \eqref{e-erase}.
Even for this case, our second main result \eqref{true} is still valid and claims that when some devices can perform the state transition \eqref{e-erase} within the error $\delta$ for arbitrary $\alpha$, the device must contain much coherence in proportion to $1/\delta$.

In the application of our result, we set the two qubits as the system $S$, and an external system $E$ as the implementation device.
We assume that the magnetization $A_S=(\ket{1}\bra{1}-\ket{0}\bra{0})^{\otimes2}$ is conserved in the entire system.
That is, we assume that an implementation set $\calI=(\calH_E,A_E,\rho_E,U_{SE})$ realizes the state transition \eqref{e-erase} within the error $\delta$ for arbitrary $\alpha$, and that the total dynamics $U_{SE}$ satisfies $[U_{SE},A_S+A_E]=0$.
Subsequently, three inequalities $\delta(\ket{00})\le\delta$, $\delta(\ket{11})\le\delta$ and $\delta(\alpha\ket{00}+\beta\ket{11})\le\delta$ are satisfied by setting the specific unitary transformation $U'_S:=\ket{00}\bra{00}+\ket{01}\bra{01}+\ket{11}\bra{10}+\ket{10}\bra{11}$.
Note that the states $\ket{00}$ and $\ket{11}$ are the eigenstates of $A''_S-A_S$ with $A''_S=U'^{\dagger}_SA_SU'_S$.
Substituting the above in \eqref{true}, the initial system $\rho_E$ of $E$ satisfies
\begin{align}
\sqrt{\calF(\rho_E)}\ge&\max_{\alpha}\frac{\chi(\alpha\ket{00}+\beta\ket{11},\{\ket{00},\ket{11}\})}{5\sqrt{2}\delta}-4\|A_S\| \nt \\
\ge &\frac{1}{5\sqrt{2}\delta}-4\|A_S\| . \label{costfore-erase}
\end{align}
In the second inequality, we used the fact that the maximum value of $\chi(\alpha\ket{00}+\beta\ket{11},\{\ket{00},\ket{11}\})$ is 1.
In this derivation, we do \textit{not} assume that the dynamics given by $\calI$ approximates to $U'_S$ the initial states other than $\alpha\ket{00}+\beta\ket{11}$.
Even when the CPTP-map given by $\calI$ is far from unitary, the inequality \eqref{costfore-erase} holds.
The inequality \eqref{costfore-erase} demonstrates that a large coherence is required for entanglement erasure, even considering the implementation of gates that are not unitary gates.

\section{extension to framework of $G$-covariant operations}\label{sectionGcov}
In this section, we extend our framework to the resource theory of asymmetry.
As a consequence of this extension, we show that our results also clarify the amount of necessary and sufficient resource to implement non-free unitary, in the case of resource theory of $U(1)$-asymmetry.

At first, we introduce the framework of resource theory of asymmetry.
Our framework is the standard one used in Refs. \cite{Bartlett, Gour, Marvian, Marvian2016, Marvian-thesis,Takagi2018, Marvian2018, Lostaglio2018}.
In the resource theory of asymmetry, the free operations are given as $G$-covariant operations that are symmetric with respect to some symmetry group $G$.
To be concrete, the $G$-covariant operation is the quantum operation $\calE$ satisfying the following equation for the unitary representation of the group $\{U_g\}_{g\in G}$ \cite{footnote-asymmetry}:
\begin{align}
{\cal E}(U_{g}(...)U^{\dagger}_{g})=U_{g}{\cal E}(...)U^{\dagger}_{g},\enskip \forall g\in G.\label{G-cov-def}
\end{align}
Also, the free states are given as invariant states with respect to the transformation by $\{U_g\}_{g\in G}$:
\begin{align}
\rho=U_g\rho U^{\dagger}_g,\enskip \forall g\in G.
\end{align}
The above free operations (G-covariant operations) the free states (G-invariant states) satisfy the following important properties:
\begin{itemize}
\item[P1]{Every $G$-covariant operation can be realized by proper set of free state and unitary $U_g$.
In fact, for an arbitrary $G$-covariant operation ${\cal E}$ on a quantum system $A$, there is another system $B$ that has a unitary representation $\{U_g\}_{g\in G}$ of $G$ on $AB$, and we can realize ${\cal E}$ with a free state $\rho$ on $B$ and a unitary $U_{g}$ in $\{U_g\}_{g\in G}$ as follows \cite{Keyl,Marvian-thesis}:
\begin{align}
{\cal E}(...)=\Tr_{B}[U_g(...\otimes\rho)U^{\dagger}_g].
\end{align}
}
\item[P2]{We cannot transform a free state to non-free state by $G$-covariant operation. Namely, if a state $\rho$ is $G$-invariant and a quantum operation ${\cal E}$ is $G$-covariant, then the state ${\cal E}(\rho)$ is also $G$-invariant.}
\end{itemize}

Property P2 naturally leads us to the notion that there is a kind of ``resource'' that does not increase under $G$-covariant operations.
The resource shows the degree of how far a non-free state is from free states.
There are many researches about how to measure the amount of the resource in the resource theory of asymmetry \cite{Bartlett, Gour, Marvian, Marvian2016, Marvian-thesis,Takagi2018}, and they have shown that the following properties are desirable for good measures of asymmetry. 
\begin{itemize}
\item{The measure $R$ does not increase through the $G$-covariant operations.}
\item{The measure $R$ is non-negative, and is zero if and only if $\rho$ is $G$-invariant.}
\end{itemize}
The quantum Fisher information used is one of well-known measures of asymmetry satisfying the above properties \cite{Marvian-thesis,Takagi2018}.
(It also satisfies many other desirable properties including the additivity for the product states.)

Now, we have introduced the framework of resource theory of asymmetry.
It is natural thought to consider the amount of necessary and sufficient resource to implement $G$-incovariant operation by $G$-covariant operation. 
This is the key problem of resource theory of quantum channels in case of resource theory of asymmetry.
Here, we extend our framework to $G$-covariant operations, and show that our results give a complete answer to this question for the case where the implemented operation is unitary.
We also partially answer to the question for the case of non-unitary operations.

We focus on the case of $G=U(1)$.
In this case, the unitary representation $\{U_{g}\}_{g\in G}$ satisfies the following equation for some Hermitian operator $A$:
\begin{align}
U_{g}=e^{-igA}.
\end{align}
Due to this fact and Property P1, we can easily show that our framework in the main text is equivalent to the simulation of unitary operation $U_S$ under $U(1)$-covariant operation.
In the extension, the implementation set becomes ${\cal K}:=({\cal H}_E,A_E,\rho_E,{\cal E}_{SE})$, where ${\cal E}_{SE}$ is a $U(1)$-covariant operation for $\{U_{g}\}_{g\in G}$ such that $U_g=e^{-ig(A_S+A_E)}$.
Then, for the desired unitary $U_S$ on $S$, we can define the implementation error $\delta_{{\cal K}}$ and coherence cost ${\cal F}'_{U_S,\delta}$ in the same way as \eqref{implementation-error} and \eqref{deff}.
Due to Property P1, the additivity of the quantum Fisher information for the product states, and the fact that the quantum Fisher information is zero for the G-invariant state, Theorem \ref{T1} immediately gives
\begin{align}
\sqrt{{\cal F}'_{U_S,\delta}}\ge\frac{{\cal A}_{U_S}}{\delta}-4\|A_S\|.
\end{align}
Also, because $U_g$ is $G$-covariant operation, Theorem \ref{T2} immediately gives
\begin{align}
\sqrt{{\cal F}'_{U_S,\delta}}\le\frac{{\cal A}_{U_S}}{\delta}+\sqrt{2}\|A_S\|.
\end{align}
Hence, we obtain 
\begin{align}
\sqrt{{\cal F}'_{U_S,\delta}}=\frac{{\cal A}_{U_S}}{\delta}+O(\|A_S\|),\enskip \delta\rightarrow0.
\end{align}
Therefore, our results clarify the amount of necessary and sufficient resource to implement non-free unitary, in the case of resource theory of $U(1)$-asymmetry.

In the same way as the above discussion, we can show that the inequality \eqref{true} also holds for the implementation set $\calK$. 
It give a lower bound for necessary coherence to implement the non-unitary G-incovariant operation $\calE_S$ under G-covariant operations, in the case where the operation $\calE_S$ is close to unitary for a partial space of the Hilbert space of the target system $S$.

\section{Derivation of lower bounds of coherence cost}
\subsection{Main idea of proof of lower bounds of coherence cost}

Before discussing the proofs of the lower bounds \eqref{T1-2} and \eqref{true}, we will present the main idea (outline) of these proofs.
The key
% KS observation
ingredient
in these proofs is the following:

\begin{lemma}\lb{L3}
Consider two quantum states, $\sigma_1$ and $\sigma_2$, and an observable $X$.
We define the difference between the expectation values of $X$ for $\sigma_1$ and $\sigma_2$ as $\Delta :=|\Tr[X(\sigma_1-\sigma_2)]|$, and the Bures distance
\eq{
L(\sigma_1, \sigma_2):=\sqrt{2(1-\Tr [\sqrt{\sqrt{\sigma_1}\sigma_2\sqrt{\sigma_1}}])}.
}
We obtain the following key relation for the case $L(\sigma_1,\sigma_2)\le1$, which was first presented in Ref.~\cite{TSS}:
\eqa{
\Delta \leq \ell(\sigma_1, \sigma_2) (V_{X}(\sigma_1)+V_{X}(\sigma_2)),
}{L3-1}
where we defined $\ell$ as
\eq{
\ell (\sigma_1, \sigma_2):= \frac{L(\sigma_1, \sigma_2)}{1-L(\sigma_1, \sigma_2)}.
} 
\end{lemma}
The quantity $\ell$ becomes small when two states $\sigma_1$ and $\sigma_2$ are close to each other, and $V_X$ is the standard deviation of $X$.
The key relation \eqref{L3-1} claims that the expectation values of $X$ can differ significantly in two states only when (i) these two states differ significantly, or (ii) at least one of the standard deviations of $X$ in these states is large.
To understand the meaning of the condition (ii), we provide an example:
Consider $\braket{0|X|0}=0$, $\braket{x|X|x}=x$, and set $\ket{\sigma_1}=\ket{0}$ and $\ket{\sigma_2}=\sqrt{1-\ep}\ket{0}+\sqrt{\ep}\ket{x}$.
The difference between these two states, $\ell(\sigma_1,\sigma_2)$ or $L(\sigma_1,\sigma_2)$, depends only on $\ep$, not on $x$.
Hence, even when $\ep$ is small (i.e., two states are close to each other), $\Delta$ can increase by setting $x$ large.
In this case, the standard deviation of $X$ in $\sigma_2$ increases with $x$.

\bigskip

In the derivation of the lower bounds for coherence, we use the key relation \eqref{L3-1} by setting $X$ as $A_E$, the conserved quantity in $E$, and $\sigma_1$ and $\sigma_2$ as the two final states of $E$ with different initial states of $S$.
In addition, we use the following three relationships:
\be{
\renewcommand{\labelenumi}{(\alph{enumi})}
\item If the time evolution of $S$ is well approximated by a unitary operation, then the final states of $E$ with different initial states of $S$ are close to each other.
\item If the variance of $A_E$ for the final states of $E$ is large, then that for the initial state of $E$ is also large.
\item If the fluctuation of the exchange of the conserved quantity $A$ between $S$ and $E$ is large, then the expectation values of $A_E$ for the final states of $E$ largely varies depending on the initial states of $S$.
}
The relationship (a) is given as a consequence of the fact that very small correlation between $S$ and $E$ is formed when the time evolution of $S$ is close to unitary.
The relationships (b) and (c) are given as the consequences of the conservation law $[U_{SE},A_S+A_E]=0$.
In deriving our first main result \eqref{T1-2}, the relationship (a) connects $L(\sigma_1,\sigma_2)$ and $\delta$, (b) connects $V_X$ and $\calF(\rho_E)$, and (c) connects $\Delta$ and $\calA_{U_S}$.

\subsection{Proof of \eqref{T1-2}}

In this subsection, we demonstrate the proof of \eqref{T1-2}.
We first describe the setup in consideration and introduce some symbols.
We consider an implementation set $\calI=({\cal H}_E, A_E, \rho_E, U_{SE})$ for a unitary operation $U_{S}$.
We prepare three initial states of $S$; $\rho_{S,0}$, and $\rho_{S,1}$ and $\rho_{S,0+1}:=(\rho_{S,0}+\rho_{S,1})/2$.
We write the final state of $E$ in actual dynamics with the initial state $\rho_{S,i}$ ($i=0,1, 0+1$) as
\begin{align}
\sigma_{E,i}&:=\Tr_{S}[U_{SE}(\rho_{S,i}\otimes\rho_{E})U^{\dagger}_{SE}].
\end{align}
We also consider two special initial states of $S$ labeled as $\rho_{S,\uparrow}$ and $\rho_{S,\downarrow}$ that maximizes and minimizes the loss of the quantity $A$ in the system, respectively, through the unitary dynamics $U_S$:
\begin{equation}
\begin{split}
\rho_{S,\uparrow}
&:= \mbox{argmax}_{\rho_S} {\rm Tr}\left[\rho_S(A'_S-A_S)\right], \\
\rho_{S,\downarrow} &:= \mbox{argmin}_{\rho_S }{\rm Tr}\left[\rho_S(A'_S-A_S)\right] \, ,
\end{split}
\label{maxmin}
\end{equation}
where we used the abbreviation $A'_S=U^{\dagger}_SA_SU_S$ again.
We write the corresponding final states of $E$ as $\sigma_{E,\uparrow}$ and $\sigma_{E,\downarrow}$.

We now state the aforementioned three relationships in a concrete form.
First, the relationship (a) is represented by the following inequality:
\begin{align}
L(\sigma_{E,0},\sigma_{E,1})\le2\sqrt{2}\delta(\rho_{S,0+1}). \label{LA-1}
\end{align}
For $\delta\le1/2\sqrt{2}$, a stronger inequality
\eqa{
L(\sigma_{E,0},\sigma_{E,1})\le2\delta(\rho_{S,0+1})
}{LA-1-stronger}
is satisfied.
These inequalities indicate a clear connection between the distance of two final state in $E$ and the accuracy of implementation (for the initial state $\rho_{S,0+1}$).
We note that these inequalities apply even when $U_{SE}$ does not commute with $A_S+A_E$, and even when the dynamics of $SE$ is not unitary.
We prove the generalized version of these inequalities in Appendix \ref{AppendixA}.

\begin{figure}
\includegraphics[width=7.5cm]{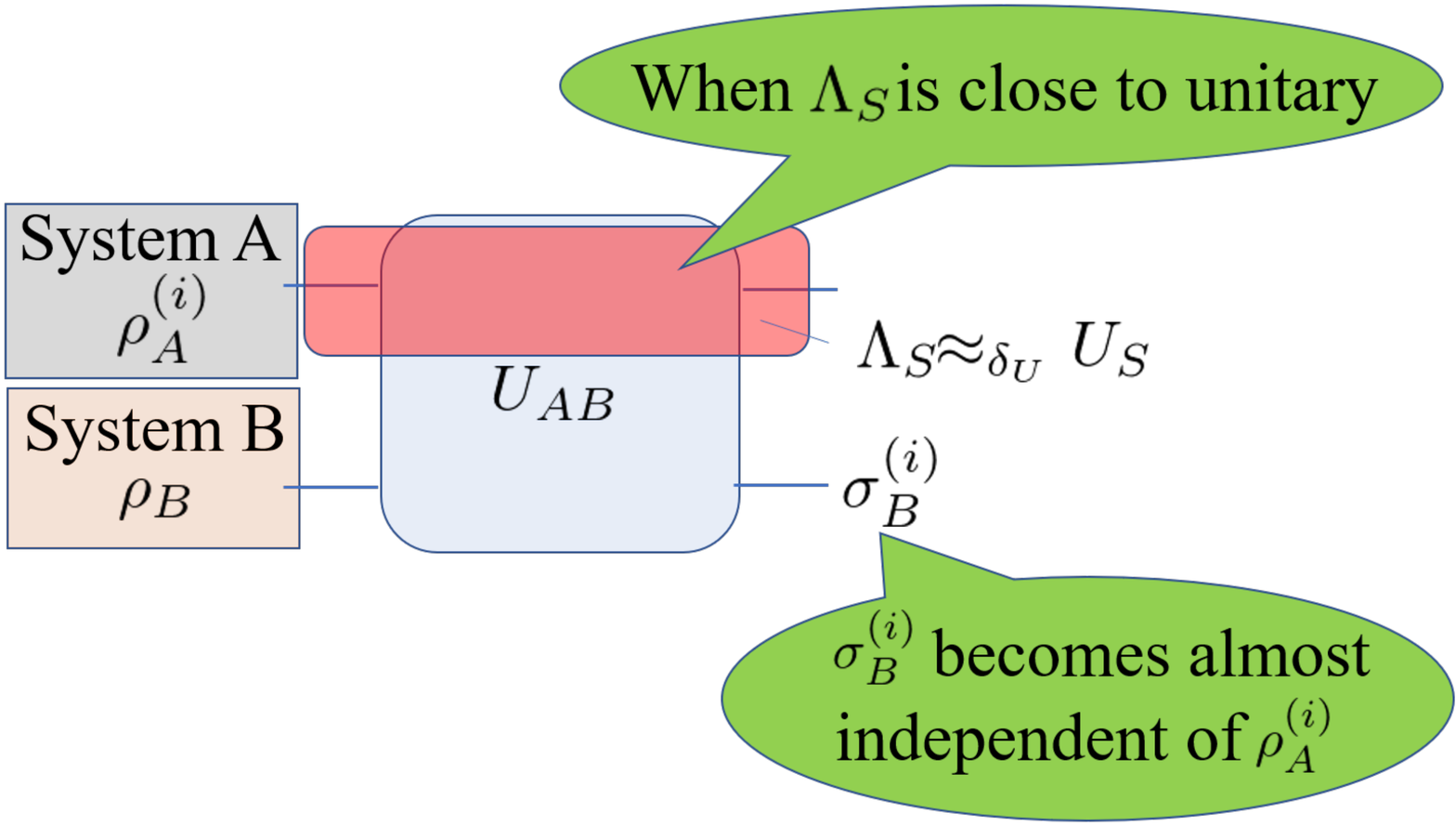}
\caption{
Schematic diagram of the inequality \eqref{LA-1}.
When the dynamics of the system $A$ is close to unitary, the final state of $B$ is close to independent of the initial state of $A$.}
\label{nocorrelation}
\end{figure}

Next, the relationship (b) is represented by the following inequality:
\eq{
V_{A_E}(\sigma_{E,\uparrow})+V_{A_E}(\sigma_{E,\downarrow})\le2(V_{A_E}(\rho_E)+\|A_S\|). \label{help1}
}
The term $\| A_S\|$ is a correction term.
This inequality connects the variance in the final state and that in the initial state.

Finally, the relationship (c) is represented by the following inequality:
\eq{
2\calA_{U_S}\le\Delta+4\delta(\rho_{S,\uparrow+\downarrow})\|A_S\|, \label{help2}
}
where we set $\Delta:=|\Tr[(\sigma_{E,\uparrow}-\sigma_{E,\downarrow})A_E]|$.
Again $\delta(\rho_{S,\uparrow+\downarrow})\|A_S\|$ is a correction term.
This inequality connects the degree of violation of the conservation of $A$ and the difference between the expected change in $A_S$ with the initial state $\rho_{S,\up}$ and $\rho_{S,\down}$.
We prove these two inequalities in Appendix \ref{s:proof-help}.

\begin{proofof}{Theorem \ref{T1}}
To prove \eqref{T1}, it suffices to show the following inequality
\begin{align}
\sqrt{\calF(\rho_E)}\ge\frac{\calA_{U_S}}{\delta(\rho_{S,\uparrow+\downarrow})}-4\|A_S\|.\label{T1-2'}
\end{align}
for any implementation set $({\cal H}_E,A_E,\rho_E,U_{AE})$ which implements $U_S$ within error $\delta$.
We divide the problem into two cases: $\delta >\calA_{U_S}/4\|A_S\|$ and $\delta \le\calA_{U_S}/4\|A_S\|$.
The former is trivial because in this case the right-hand side of \eqref{T1-2'} is negative while the quantum Fisher information $\sqrt{\calF}$ is always nonnegative.
In the following, we consider the latter case:$\delta \le\calA_{U_S}/4\|A_S\|$.

We first show \eqref{T1-2'} in the case that $\rho_E$ is a pure state. 
Since $\delta \le\calA_{U_S}/4\|A_S\|\leq 1/4$, the stronger inequality \eqref{LA-1-stronger} is satisfied, which suggests
\begin{align}
L(\sigma_{E,\uparrow},\sigma_{E,\downarrow}) \le 2\delta(\rho_{S,\uparrow+\downarrow}).
\end{align}
Substituting the above relation, \eqref{help1}, and \eqref{help2}, to Lemma \ref{L3}
\begin{align}
\Delta\le L(\sigma_{E,\uparrow},\sigma_{E,\downarrow})(V_{A_E}(\sigma_{E,\uparrow})+V_{A_E}(\sigma_{E,\downarrow})+\Delta)
\end{align}
and using $\Delta\le2\|A_S\|$, we obtain
\begin{align}
\calA_{U_S}\le\delta(\rho_{S,\uparrow+\downarrow})(2V_{A_E}(\rho_E)+4\|A_S\|).
\end{align}
By definition of the quantum Fisher information \eqref{F=Q}, $2V_{A_E}(\rho_E)=\sqrt{\calF(\rho_E)}$ holds for a pure state $\rho$.
Thus, we obtain \eqref{T1-2'} for the case that $\rho_E$ is pure.

Next, we show \eqref{T1-2'} in the case that $\rho_E$ is a mixed state.
We expand the initial state of the external system as $\rho_E:= \sum_j p_j \phi_{E,j}$ satisfying $\calF(\rho_E)=4\sum_{j}p_{j}V^{2}_{A_E}(\phi_{E,j})$.
We denote $\delta(\rho_{S,\uparrow+\downarrow})$ for the case that the initial state of $E$ is $\phi_{E,j}$ by $\delta_{j}$:
\eq{
\delta_j:=L_e(\rho_{S,\uparrow+\downarrow},\Lambda_{U_S^\dagger}\circ\Lambda_{S,j}),
}
where$\Lambda_{S,j}$ is the dynamics of $S$ for the case that $\phi_j$ is the initial state of $E$, i.e., $\Lambda_{S,j}(...):=\Tr_{E}[U_{SE}(...\otimes \phi_{E,j})U^{\dagger}_{SE}]$.
The inequality \eqref{T1-2'} for pure states, which we have already proven, yields
\begin{align}
2V_{A_E}(\phi_{E,j})&\ge\frac{\calA_{U_S}}{\delta_{j}}-4\|A_S\| \label{T1-2''}
\end{align}
for any $j$.
Here, let us define $k(x):=(\max\{0,\frac{\calA_{U_S}}{x}-4\|A_S\|\})^2$.
Due to \eqref{T1-2''} and the downward convexity of $k(x)$, we have
\begin{align}
\calF(\rho_E)=4\sum_{j}p_{j}V_{A_E}(\phi_{E,j})^2\ge&\sum_{j}p_{j}k(\delta_{j}) 
\ge k(\sum_{j}p_{j}\delta_j).\label{32}
\end{align}
Hence, to prove \eqref{T1-2'} for a mixed state, it suffices to show
\begin{align}
\sum_{j}p_{j}\delta_{j}\le\delta(\rho_{S,\uparrow+\downarrow}), \label{lld}
\end{align}
because the function $k$ is non-increasing.

Finally, we shall show \eqref{lld}.
We employ the following equality:
\begin{align}
&\left(1-\frac{\delta(\rho_{S,\uparrow+\downarrow})^2}{2}\right)^2\nonumber\\
&=\bra{\psi_{SR,\uparrow+\downarrow}}U^{\dagger}_{S}\Lambda_{S}(\psi_{SR,\uparrow+\downarrow})U_S\ket{\psi_{SR,\uparrow+\downarrow}}\nonumber\\
&=\sum_{j}p_{j}\bra{\psi_{SR,\uparrow+\downarrow}}U^{\dagger}_{S}\Lambda_{S,j}(\psi_{SR,\uparrow+\downarrow})U_S\ket{\psi_{SR,\uparrow+\downarrow}}\nonumber\\
&=\sum_{j}p_{j}\left(1-\frac{\delta^{2}_{j}}{2}\right)^2,\label{43}
\end{align}
where $\psi_{SR,\uparrow+\downarrow}$ is the purification of $\rho_{S,\uparrow+\downarrow}$.

The tangent line passing through the point $(x,y)=(\sqrt{2},0)$ and touching curve $y=g(x):=(1-{x^2}/{2})^2$ from above also passes the point $(x,y)=(\frac{\sqrt{2}}{3},\frac{64}{81})$.
Therefore, for any probability distribution $\{q_j\}$ and real numbers $0\le x_j\le\sqrt{2}$, the following inequality holds
\begin{align}
\sum_{j}q_jg(x_j)\le g'(\sum_{j}q_jx_j),\label{44}
\end{align}
where $g'$ is an upward convex function defined as
\begin{equation}
g'(x):=
\left\{
\begin{array}{ll}
g(x) & (0\le x \le \frac{\sqrt{2}}{3}) \\
\frac{16\sqrt{2}}{27}(\sqrt{2}-x) & (\frac{\sqrt{2}}{3}<x\le\sqrt{2}).\label{45}
\end{array}
\right.
\end{equation}
Due to \eqref{43}, \eqref{44} and \eqref{45}, 
\begin{align}
g'(\delta(\rho_{S,\uparrow+\downarrow}))=g(\delta(\rho_{S,\uparrow+\downarrow}))\le g'(\sum_{j}p_j\delta_j).
\end{align}
Here we use $\delta(\rho_{S,\uparrow+\downarrow})\le\delta\le1/4$.
Since $g'$ is a non-increasing function of $x$, we obtain \eqref{lld}.
\end{proofof}

\subsection{The proof of \eqref{true} for pure states}

We next consider the inequality for a single initial state \eqref{true}.
Since the complete proof is a little complicated and we need many additional treatment for some correction terms, we here only prove it for pure states $\rho_S$ and $\rho_E$ with using several inequalities that are shown in Appendix.
We shall present a complete proof in Appendix.~\ref{s:pf-true}.

We first introduce some symbols used in the proof.
We denote the desired final state and realized final state by
\begin{align}
\ket{\rho'_S}&:=U_S\ket{\rho_S},\\
\ket{\Psi'_{SE}}&:=U_{SE}\ket{\rho_{S}}\ket{\rho_E},\\
\sigma_{E}&:=\Tr_{S}[\Psi'_{SE}], \\
\sigma_{S}&:=\Tr_{E}[\Psi'_{SE}].
\end{align}
In a similar manner, the final state of $\psi_i$ is denoted by
\begin{align}
\ket{\psi'_{i}}&:=U_S\ket{\psi_{i}}\\
\ket{\Psi'_{i,SE}}&:=U_{SE}\ket{\psi_{i}}\ket{\rho_E},\\
\sigma_{i,E}&:=\Tr_{S}[\Psi'_{i,SE}], \\
\sigma_{i,S}&:=\Tr_{E}[\Psi'_{i,SE}].
\end{align}
With noting the definition of $\delta (\rho)$, Uhlmann's theorem~\cite{hayashi} tells that there are two pure states $\phi'_{E}$ and $\phi'_{i,E}$ satisfying
\begin{align}
|\braket{\Psi'_{SE}|\rho'_S\otimes \phi'_{E}}|&=1-\frac{\delta (\rho_S)^2}{2}\label{51}\\
|\braket{\Psi'_{i,SE}|\psi'_{i}\otimes \phi'_{i,E}}|&=1-\frac{\delta (\psi_{i})^2}{2},
\end{align}
where we wrote $\ket{\rho}\otimes \ket{\phi}$ as $\ket{\rho \otimes \phi}$.
We also employ the abbreviation:
\eq{
\overline{\delta}^2(\rho_S, \{ \psi_i\} ):=\delta_{}(\rho_S)^2+\sum_ir_i\delta_{}(\psi_i)^2.
}
Since \eqref{true} becomes trivial when $\overline{\delta}>\frac{\chi(\rho_s,\{\psi_i\})}{20\|A_S\|}$, we hereafter focus on the case of $\overline{\delta}\le\frac{\chi(\rho_s,\{\psi_i\})}{20\|A_S\|}$.

Similar to \eqref{T1-2}, the derivation of \eqref{true} is based on the relations corresponding to Lemma.\ref{L3} and the relationship (a)-(c).
Let us start from the relationship (c).
The \eqref{true}-version relationship (c) is represented as
\balign{
\chi(\rho_S,\{\psi_i\})\le\sqrt{\sum_{i} r_i \Delta_i ^2}+\Delta'+4\overline{\delta}(\rho_S)\|A_S\|, \lb{c-single}
}
where we defined
\balign{
\Delta_i&:=|\left<A_E\right>_{\sigma_{i,E}}-\left<A_E\right>_{\phi'_E}|,\nonumber\\
\Delta'&:=|\left<A_E\right>_{\sigma_{E}}-\left<A_E\right>_{\phi'_E}|.
}
To bound $\Delta_i$ and $\Delta'$, we use Lemma.~\ref{L3}.
With using Lemma \ref{L3} and \eqref{51}, $\Delta'$ is bounded as follows:
\begin{align}
\Delta'\le\frac{\delta(\rho_S)}{1-\delta(\rho_S)}(V_{A_E}(\sigma_E)+V_{A_E}(\phi'_E)).
\end{align}
(We can apply Lemma \ref{L3} to this situation because of $\delta(\rho_S)\le\overline{\delta}(\rho_S,\{\psi_i\})\le\frac{\chi(\rho_S,\{\psi_i\})}{20\|A_S\|}\le\frac{1}{10}$)
We also bound $\sum_ir_i\Delta^2_i$ with using Lemma.~\ref{L3} in the following form:
\balign{
\sum_ir_i\Delta_i^2
\le&\sum_ir_iL(\sigma_{i,E},\phi'_E)^2(V_{A_E}(\sigma_{i,E})+V_{A_E}(\phi'_{E})+\Delta_i)^2. \lb{L3-single}
}
The third term $\Delta_i$ in the right hand side of \eqref{L3-single} is bounded by $\Delta'$ and $\|A_S\|$ as follows (proof is in Appendix \ref{s:pf-true}):
\balign{
\Delta_i\le\Delta'+2\|A_S\|.
}
We note that $V_{A_E}(\rho_E)$ has a direct connection to the Fisher information $\calF(\rho_E)$ in case with a pure state.
Therefore, we obtain \eqref{true} by bounding the three remaining terms $\sum_ir_iL(\sigma_{i,E},\phi'_E)^2$, $V_{A_E}(\sigma_{i,E})$ and $V_{A_E}(\phi'_{E})$ by $V_{A_E}(\rho_E)$ and $\overline{\delta}(\rho_S,\{\psi_i\})$.
To do so, we use the relationships (a)-(c).

The \eqref{true}-version relationship (a) is represented as
\eqa{
\sum_{i}r_i L(\phi'_{E},\sigma_{i,E})^2\le 4\overline{\delta}(\rho_S, \{ \psi_i\} )^2,
}{a-single}
which is close to the relation \eqref{LA-1}.
The \eqref{true}-version relationship (b) is represented as
\balign{
V_{A_E}(\sigma_{i,E})&\leq V_{A_E}(\rho_E)+\|A_S\|,\\
V_{A_E}(\phi'_{E})&\leq \frac{\sqrt{\frac{1}{4}\|A_S\|^2+V_{A_E}(\rho_E)^2}}{1-\frac{\delta^2(\rho_S)}{2}}\label{b-single}
}
which are close to the relation \eqref{help1}.
These relations are shown in the similar manner to those for \eqref{LA-1} and \eqref{help1}, and shown in Appendix \ref{s:pf-true}.
Combining these relations and evaluating all correction terms, we arrive at the desired relation \eqref{true}.

\section{\label{sec:theot2}Derivation of sufficient condition of coherence cost (Theorem \ref{T2})}

We prove Theorem \ref{T2} by using the following lemma:
\begin{lemma}\label{CL3}
We take an one-dimensional continuous system as ${\cal H}_E$, and set the position operator $x$ on it as the Hermitian $A_E$.
Given an arbitrary unitary $U_S:=\sum_{ij}u_{ij}\ket{i}\bra{j}$ on $S$ ($\ket{i}$ and $\ket{j}$ are the eigenvectors of $A_S$ whose eigenvalues are $h_i$ and $h_j$), and an arbitrary positive real number $\zeta$.
We define $U_{SE}$ and $\phi_{\zeta}$ as follows:
\begin{align}
\ket{\phi_{\zeta}}&:=C\int^{\infty}_{-\infty}\sqrt{e^{-\frac{x^{2}}{2\zeta^{2}}}}\ket{x}dx\\
U_{SE}&:=\sum_{ij}u_{ij}\ket{i}\bra{j}\otimes e^{-ip(h_i-h_j)}, \label{CL3-2}
\end{align}
where $p$ represents the momentum operator and $C$ is a normalization constant.
By construction, $[U_{SE},A_S+A_E]=0$ is satisfied.

Then, for $\zeta\geq 9\calA_{U_S}/2\sqrt{2}$, the implementation set $\calI_{\zeta}=({\cal H_E}, A_E, \phi_{\zeta}, U_{SE})$ provides a good implementation of $U_S$ in the following sense:
\begin{align}
\max_{\rho_S}L_{e}(\rho_S,\Lambda_{U^{\dagger}_S}\circ\Lambda_S)&\le\frac{\calA_{U_S}}{2\zeta}\left(1+\frac{\|A_S\|}{\sqrt{2}\zeta}\right)\label{CL3-3K}.
\end{align}
\end{lemma}

We leave the proof of Lemma \ref{CL3} to the Appendix.~\ref{s:pf-suf}.
Intuitively speaking, the Gaussian state serves as a good external system to {\it absorb} the back action of the change in $A$.

\begin{proofof}{Theorem \ref{T2}}
Given an arbitrary $U_S$ on $S$, arbitrary precision $\delta$ with $0\le \delta \le\frac{4\sqrt{2}\calA_{U_S}}{9\|A_S\|}$ and arbitrary real number $\calF$ satisfying $\sqrt{\calF}\ge\frac{\calA_{U_S}}{\delta}+\sqrt{2}\|A_S\|$ we construct an implementation set $\calI$ for $U_S$ satisfying $\delta_{\calI}\le\delta$ and $\calF(\rho_E)=\calF$.
We set a real positive number $\zeta$ as follows:
\begin{align}
2\zeta:=\sqrt{\calF} \label{C1-1}
\end{align}
We show that the implementation set $({\cal H}_{E}, A_E, \phi_{\zeta}, U_{SE})$ constructed in Lemma \ref{CL3} is $\calI$ which we seek.

The relation $\calF(\phi_{\zeta})=\calF$ is easy to obtain from \eqref{C1-1} by inserting $\sqrt{\calF(\phi_{\zeta})}=2\zeta$.
The accurate implementation
\begin{align}
\max_{\rho_S}L_{e}(\rho_S,\Lambda_{U^{\dagger}_S}\circ\Lambda_S)\le\delta,
\end{align}
is also confirmed by substituting \eqref{C1-1} to \eqref{CL3-3K}, which means that $\calI_{\zeta}$ implements $U_S$ within error $\delta$.
\end{proofof}

\section{Summary and discussion}
In this paper, we established simple relations between quantum coherence and asymmetry (violation of a conservation law).
The coherence cost to realize unitary dynamics in a partial system under a symmetry (a conservation law) in a total system is asymptotically equal to the ratio between the degree of asymmetry of the implemented unitary and the implementation error.
We derive the upper and lower bounds for the coherence cost that are asymptotically identical in the region where the error is small.
This asymptotic equation quantitatively links two fundamental concepts in physics, i.e., symmetry and coherence.

Our results are applicable even when the whole system satisfies multiple conservation laws.
If the desired unitary dynamics alters two physical quantities and if the two physical quantities are conserved in the total system, then the external system must have the coherence required by Theorem \ref{T1} for each quantity.

Since our results are valid for any unitary operation, there are various applications of our results.
In this paper, the implementation of quantum heat engine, resource theory and entanglement erasure are described as examples.
In addition to these examples, our results are applicable whenever we try to realize a time-dependent Hamiltonian or to perform some control while maintaining the quantum superposition.

Finally, we present a possible extension of our result to arbitrary CPTP-maps.
Let us consider an arbitrary CPTP map $\calE_S$ on $S$.
We will implement this CPTP map by using the same type of implementation set $\calI=({\cal H}_E, A_E, \rho_E, U_{SE})$.
Its total dynamics $U_{SE}$ conserves $A_S+A_E$ and the initial state $\rho_E$ might have coherence, i.e., $[U_{SE},A_S+A_E]$ must be zero and $\calF(\rho_E)$ can be zero.
To define the degree of asymmetry (violation of the conservation of $A_S$), we consider another type of implementation $\calJ=({\cal H}_E, A_E, \eta_E, V_{SE})$, whose initial state does not have coherence, that is, $\calF(\rho_E)$ must be zero, and $V_{SE}$ might not conserve $A_S+A_E$, that is, $[V_{SE},A_S+A_E]$ might be nonzero.
%With using these two types of implementation, we can define the degree of asymmetry and the coherence cost of $\calE_S$.
We define the degree of asymmetry of $\calE_S$ as the minimum degree of asymmetry in all possible $\calJ$ that implements $\calE_S$ with no error \cite{footnote2}:
\begin{align}
\calA_{\calE_S}:=\min_{\calJ\models_{0}\calE_S}\calA_{V_{SE}}.
\end{align}
We also define the coherence cost of $\calE_S$ as 
\begin{align}
\calF_{\calE_S,\delta}:=\min_{\calI\models_{\delta}\calE_S}\calF(\rho_E)
\end{align}
Note that when $\calE_S=U_S$, the quantities $\calA_{\Lambda_S}$ and $\calF_{\calE_S,\delta}$ reduce to $\calA_{U_S}$ and $\calF_{U_S,\delta}$, respectively. 
Hence, $\calA_{\Lambda_S}$ and $\calF_{\Lambda_S,\delta}$ are generalizations of $\calA_{U_S}$ and $\calF_{U_S,\delta}$.
Theorem \ref{T2} provides the same form of inequality with these quantities:
\begin{align}
\sqrt{\calF_{\calE_S,\delta}}\le\frac{\calA_{\calE_S}}{\delta}+\sqrt{2}\|A_S\|.
\end{align}
However, unfortunately we do not have an inequality similar to Theorem \ref{T1}.
If such an inequality is shown, we obtain the following asymptotic relation in a concise form:
\begin{align}
\mbox{Conjecture: }\sqrt{\calF_{\calE_S,\delta}}=\frac{\calA_{\calE_S}}{\delta}+O(\|A_S\|).
\end{align}
We leave this problem as a future work.

\bigskip

\acknowledgments

We thank Hiroshi Nagaoka, Tomohiro Ogawa, Satoshi Ishizaka and Eyuri Wakakuwa for the fruitful discussion and helpful comments.
The present work was supported by JSPS Grants-in-Aid for Scientific Research No. JP19K14610 (HT), No. JP19K14615 (NS), No. JP25103003 (KS), and No. JP16H02211 (KS).

\appendix

\section{Proof of \eqref{LA-1} and \eqref{LA-1-stronger}}\label{AppendixA}
In this Appendix, we prove \eqref{LA-1} and \eqref{LA-1-stronger}.
Precisely speaking, we prove the following generalized version of \eqref{LA-1} and \eqref{LA-1-stronger}:

\begin{lemma}\label{L1}
Consider two quantum systems $A$ and $B$.
Let $\Lambda_{AB}$ be a CPTP map on the composite system $AB$ and $U_A$ be a unitary operation on $A$.
We consider three possible initial states of $A$: $\rho^{(0)}_{A}$, $\rho^{(1)}_{A}$, and $\rho^{(0+1)}_{A}:=(\rho^{(0)}_{A}+\rho^{(1)}_{A})/2$.
We write the initial state of $B$ as $\rho_{B}$.
We refer to the final states of $AB$ and $B$ with the initial state $\rho^{(i)}_{A}$ ($i=0,1,0+1$) as 
\begin{align}
\sigma^{(i)}_{AB}&:=\Lambda_{AB}(\rho^{(i)}_{A}\otimes\rho_{B}),\\
\sigma^{(i)}_{B}&:=\Tr_{A}[\sigma^{(i)}_{AB}].
\end{align}

We refer to the time-evolution of $A$ determined by $\Lambda_{AB}$ and $\rho_{B}$ as $\Lambda_{A}(\rho):=\Tr_{B}[\Lambda_{AB}(\rho\otimes\rho_B)]$.
Using this symbol, we define the accuracy of implementation of $U_A$ by $\Lambda_{AB}$ for the initial states $\rho^{(i)}_{A}$ ($i=0,1,0+1$) as
\begin{align}
\delta^{(i)}_{U}:=L_{e}(\rho^{(i)}_{A},\Lambda_{U^{\dagger}_A}\circ\Lambda_{A}).
\end{align}

In this setup, we have the following results: $\\$
1. The following inequality holds:
\begin{align}
L(\sigma^{(0)}_{AB}, U_{A}\sigma^{(0)}_{A}U^{\dagger}_{A}\otimes\sigma^{(0)}_{B})\le2\delta^{(0)}_{U}.\label{L1-2}
\end{align}$\\$
2. There exists a state $\sigma'^{(0+1)}_{B}$ of $B$ such that
\begin{align}
L(\sigma^{(0)}_{B},\sigma'^{(0+1)}_{B})+L(\sigma'^{(0+1)}_{B},\sigma^{(1)}_{B})\le2\sqrt{2}\delta^{(0+1)}_{U}.\label{L1-1}
\end{align}
Moreover, if $\delta^{(0+1)}_{U}\le1/2\sqrt{2}$ holds, there exists a state $\sigma'^{(0+1)}_{B}$ of $B$ such that
\begin{align}
L(\sigma^{(0)}_{B},\sigma'^{(0+1)}_{B})+L(\sigma'^{(0+1)}_{B},\sigma^{(1)}_{B})\le2\delta^{(0+1)}_{U}.\label{L1-1-stronger}
\end{align}
If $\rho_B$ is a pure state and $\Lambda_{AB}$ is a unitary operation, the aforementioned $\sigma'^{(0+1)}_{B}$ is a pure state.
\end{lemma}

\begin{proofof}{Lemma \ref{L1}}
We first introduce some symbols.
We take the purification $\psi^{(i)}_{AR_A}$ of $\rho^{(i)}_{A}$(i=0,1,0+1) such that $\rho^{(0)}_{R_A}:=\Tr_{A}[\psi^{(0)}_{AR_A}]$ and $\rho^{(1)}_{R_A}:=\Tr_{A}[\psi^{(1)}_{AR_A}]$ are pure states and orthogonal to each other.
In that case, $\ket{\psi^{(0+1)}}_{AR}=(\ket{\psi^{(0)}}_{AR}+\ket{\psi^{(1)}}_{AR})/\sqrt{2}$ holds.
We write the purification of $\rho_B$ as $\psi_{BR_B}$.
We employ the Steinspring representation \cite{hayashi} of $\Lambda_{AB}$, that is, we describe $\Lambda_{AB}(\rho)$ by using a pure state $\psi_{C}$ and a unitary transformation $U_{ABC}$ as $\Lambda_{AB}(\rho)=\Tr_{C}[U_{ABC}(\rho\otimes\psi_{C})U^{\dagger}_{ABC}]$.
We denote the initial and final states of the total system $AR_ABR_BC$ by
\balign{
\psi^{(i)}_{tot}&:=\psi^{(i)}_{AR_A}\otimes\psi_{BR_B}\otimes\psi_C \\
\psi'^{(i)}_{tot}&:=U_{ABC}\psi^{(i)}_{tot}U^{\dagger}_{ABC},
}
respectively.
We also denote the final states of $AR_A$ and $BR_BC$ by
\balign{
{\sigma}^{(i)}_{AR_A}&:=\Tr_{BR_BC}[\psi'^{(i)}_{tot}] \\
{\sigma}^{(i)}_{BR_BC}&:=\Tr_{AR_A}[\psi'^{(i)}_{tot}],
}
respectively.

Uhlmann's theorem suggests that the definition of $\delta^{(i)}_{U}$,
\begin{align}
\delta^{(i)}_U=L(U_A\psi^{(i)}_{AR_A}U^{\dagger}_A,\sigma^{(i)}_{AR_A}),
\end{align}
has another expression with a proper pure state $\phi'^{(i)}_{BR_BC}$ as
\begin{align}
\delta^{(i)}_U=L(U_A\psi^{(i)}_{AR_A}U^{\dagger}_A\otimes\phi'^{(i)}_{BR_BC},\psi'^{(i)}_{tot}).\label{L1-A}
\end{align}
Owing to the contractivity of the Bures distance, by taking the partial trace of $AR_A$ in \eqref{L1-A} we obtain
\begin{align}
\delta^{(i)}_U\ge L(\phi'^{(i)}_{BR_BC},\sigma^{(i)}_{BR_BC}).\label{L1-B}
\end{align}

We now derive \eqref{L1-1} and \eqref{L1-2} by using \eqref{L1-B}.
We first derive \eqref{L1-2}.
We start from the following triangle inequality,
\begin{align}
&L(\psi'^{(i)}_{tot}, U_A\psi^{(i)}_{AR_A}U^{\dagger}_A\otimes\sigma^{(i)}_{BR_BC})\nonumber\\
&\le
L(\psi'^{(i)}_{tot}, U_A\psi^{(i)}_{AR_A}U^{\dagger}_A\otimes\phi'^{(i)}_{BR_BC})\nonumber\\
&+
L(U_A\psi^{(i)}_{AR_A}U^{\dagger}_A\otimes\phi'^{(i)}_{BR_BC},U_A\psi^{(i)}_{AR_A}U^{\dagger}_A\otimes\sigma^{(i)}_{BR_BC}) 
\end{align}
The first and second terms of the right-hand side is bounded by \eqref{L1-A} and \eqref{L1-B}, respectively, which yields
\begin{align}
L(\psi'^{(i)}_{tot}, U_A\psi^{(i)}_{AR_A}U^{\dagger}_A\otimes\sigma^{(i)}_{BR_BC})\le2\delta^{(i)}_U.
\end{align}
By taking the partial trace of $AR_A$ in the above inequality, we obtain the desired relation \eqref{L1-2}:
\begin{align}
L(\sigma^{(i)}_{AB}, U_A\rho^{(i)}_{A}U^{\dagger}_A\otimes\sigma^{(i)}_{B})\le2\delta^{(i)}_U.
\end{align}

Next we show \eqref{L1-1}. 
We note the following relation
\begin{align}
\sigma^{(0+1)}_{BR_BC}=\frac{\sigma^{(0)}_{BR_BC}+\sigma^{(1)}_{BR_BC}}{2},
\end{align}
which comes from a relation $\Tr_{ABR_BC}[\psi'^{(a)}_{tot}]=\rho^{(a)}_{R_A}$ for $a=0, 1$, and the fact that $\rho^{(0)}_{R_A}$ and $\rho^{(1)}_{R_A}$ are orthogonal to each other.
Then, \eqref{L1-B} implies
\begin{align}
\delta^{(0+1)}_U&\ge L(\phi'^{(0+1)}_{BR_BC},\sigma^{(0+1)}_{BR_BC})\nonumber\\
&=L\left(\phi'^{(0+1)}_{BR_BC},\frac{\sigma^{(0)}_{BR_BC}+\sigma^{(1)}_{BR_BC}}{2}\right),
\end{align}
or equivalently,
\begin{align}
F\left(\phi'^{(0+1)}_{BR_BC},\frac{\sigma^{(0)}_{BR_BC}+\sigma^{(1)}_{BR_BC}}{2}\right)\ge1-\frac{(\delta^{(0+1)}_U)^2}{2}. \label{23}
\end{align}
The left-hand side of the above inequality is transformed, with noting that $\phi'^{(0+1)}_{BR_BC}$ is a pure state, into
\begin{align}
&F\left(\phi'^{(0+1)}_{BR_BC},\frac{\sigma^{(0)}_{BR_BC}+\sigma^{(1)}_{BR_BC}}{2}\right)^2\nonumber\\
&=\bra{\phi'^{(0+1)}_{BR_BC}}\frac{\sigma^{(0)}_{BR_BC}+\sigma^{(1)}_{BR_BC}}{2}\ket{\phi'^{(0+1)}_{BR_BC}} \nt \\
&=\frac{1}{2}\sum_{i=0,1}F(\phi'^{(0+1)}_{BR_BC},\sigma^{(i)}_{BR_BC})^2.
\end{align}
Combining the above equations and the relation $(1-x^2/2)^2\geq 1-x^2$, we obtain
\begin{align}
\frac{1}{2}\sum_{i=0,1}F(\phi'^{(0+1)}_{BR_BC},\sigma^{(i)}_{BR_BC})^2\ge1-(\delta^{(0+1)}_U)^2,
\end{align}
which can be evaluated as
\begin{align}
(\delta^{(0+1)}_U)^2&\ge1-\frac{1}{2}\sum_{i=0,1}F(\phi'^{(0+1)}_{BR_BC},\sigma^{(i)}_{BR_BC})^2\nonumber\\
&\ge1-\frac{1}{2}\sum_{i=0,1}F(\phi'^{(0+1)}_{BR_BC},\sigma^{(i)}_{BR_BC})\nonumber\\
&=\frac{1}{4}\sum_{i=0,1}L(\phi'^{(0+1)}_{BR_BC},\sigma^{(i)}_{BR_BC})^2\nonumber\\
&\ge \frac{1}{8}\left(\sum_{i=0,1}L(\phi'^{(0+1)}_{BR_BC},\sigma^{(i)}_{BR_BC})\right)^2\nt \\
&\ge \frac{1}{8}\left(\sum_{i=0,1}L(\sigma'^{(0+1)}_{B},\sigma^{(i)}_{B})\right)^2.
\end{align}
This is equivalent to the desired relation \eqref{L1-1}.
Here, we defined $\sigma'^{(0+1)}_{B}:=\Tr_{R_BC}[\phi'^{(0+1)}_{BR_BC}]$, and we used the relation $X^2+Y^2\ge \frac{(X+Y)^2}{2}$ for positive numbers $X$ and $Y$ in the 4th line, and the contractivity of the Bures distance in the last line.
We remark that if $\rho_B$ is pure and $\Lambda_{AB}$ is unitary, by following the above derivation without $R_BC$, we obtain the fact that $\sigma'^{(0+1)}_{B}$ is a pure state.

We finally derive \eqref{L1-1-stronger} when $\delta^{(0+1)}_{U}\le1/2\sqrt{2}$ holds.
Using again the fact that $\phi'^{(0+1)}_{BR_BC}$ is a pure state, \eqref{23} reads
\begin{align}
\frac{1}{2}\sum_{i=0,1} \(1-\frac{L(\phi'^{(0+1)}_{BR_BC},\sigma^{(i)}_{BR_BC})^2}{2}\) ^2\ge\( 1-\frac{(\delta^{(0+1)}_U)^2}{2}\) ^2
\end{align}
With noting the following relation
\begin{align}
\frac{1}{2}((1-\frac{X^2}{2})^2+(1-\frac{Y^2}{2})^2)\le(1-\frac{(X+Y)^2}{8})^2 \lb{Ap-A-support}
\end{align}
for real numbers $X$ and $Y$ satisfying $0\le X+Y\le 1$, we arrive at
\begin{align}
1-\frac{\left(\sum_{i=0,1}L(\phi'^{(0+1)}_{BR_BC},\sigma^{(i)}_{BR_BC})\right)^2}{8} \ge 1-\frac{(\delta^{(0+1)}_U)^2}{2}.
\end{align}
Here we used the relation $L(\phi'^{(0+1)}_{BR_BC},\sigma^{(0)}_{BR_BC})+L(\phi'^{(0+1)}_{BR_BC},\sigma^{(1)}_{BR_BC})\leq 2\sqrt{2}\delta_U^{(0+1)}\leq 1$, which follows from \eqref{L1-1}, in application of \eqref{Ap-A-support}.
By taking the partial trace of $R_BC$, the above inequality directly implies the desired relation \eqref{L1-1-stronger}.
%\begin{align}L(\phi'^{(0+1)}_{B},\sigma^{(0)}_{B})+L(\phi'^{(0+1)}_{B},\sigma^{(1)}_{B})\le2\delta^{(0+1)}_U.\end{align}
\end{proofof}

\section{Proof of \eqref{help1} and \eqref{help2}}\lb{s:proof-help}

\begin{proofof}{\eqref{help1}}
The conservation of $A_S+A_E$ under $U_{SE}$ yields
\begin{align}
V^{2}_{A_S}(\rho_{S,i})+V^{2}_{A_E}(\rho_{E})=&V^{2}_{A_S}(\sigma_{S,i})+V^{2}_{A_E}(\sigma_{E,i})\nonumber\\
&+2{\rm Cov}_{A_S+A_E}(U_{SE}(\rho_{S,i}\otimes\rho_{E})U^{\dagger}_{SE}),\label{B1}
\end{align}
where $i\in \{\uparrow,\downarrow\}$.
$V_{A_S}(\rho)$ represents the standard deviation of the quantity $A_S$ in $\rho$, and ${\rm Cov}_{A_S+A_E}(\sigma)$ is the covariance between $A_S$ and $A_E$ with the state $\sigma$.
Using a basic property of covariance $-V_{A_S}(\sigma_{S,i})V_{A_E}(\sigma_{E,i})\le {\rm Cov}_{A_S+A_E}(U_{SE}(\rho_{S,i}\otimes\rho_{E})U^{\dagger}_{SE})$, we arrive at
\begin{align}
&V_{A_E}(\sigma_{E,i})-V_{A_S}(\sigma_{S,i}) \nonumber\\
\le& \sqrt{V^{2}_{A_S}(\sigma_{S,i})+V^{2}_{A_E}(\sigma_{E,i})-2V_{A_S}(\sigma_{S,i})V_{A_E}(\sigma_{E,i})} \nonumber\\
\le & \sqrt{V^2_{A_E}(\rho_{E})+V^2_{A_S}(\rho_{S,i})} \nt \\
\le & V_{A_E}(\sigma_{E})+V_{A_S}(\rho_{S,i}).
\end{align}
Using a relation $\delta_{A_S}(\rho)\leq \|A_S\|_{}/2$ for any state $\rho$ and taking the sum of $i\in \{\uparrow,\downarrow\}$, we obtain \eqref{help1}.
\end{proofof}

\begin{proofof}{\eqref{help2}}
By introducing the following quantities
\begin{align}
\Delta_{\uparrow}:=&\Tr[A_E(\sigma_{E,\uparrow}-\rho_{E})]\nonumber\\
=&\Tr[A_S(\rho_{S,\uparrow}-\sigma_{S,\uparrow})],\label{B3}\\
\Delta_{\downarrow}:=&\Tr[A_E(\sigma_{E,\downarrow}-\rho_{E})]\nonumber\\
=&\Tr[A_S(\rho_{S,\downarrow}-\rho'_{S,\downarrow})],\label{B4}\\
\Delta_{U,\uparrow}:=&\Tr[A_S(\rho_{S,\uparrow}-U_{S}\rho_{S,\uparrow}U^{\dagger}_{S})],\\
\Delta_{U,\downarrow}:=&\Tr[A_S(\rho_{S,\downarrow}-U_{S}\rho_{S,\downarrow}U^{\dagger}_{S})],
\end{align}
three quantities appearing in \eqref{help2} can be written or evaluated in terms of the above quantities
\begin{align}
\Delta&=|\Delta_{\downarrow}-\Delta_{\uparrow}|,\label{8.22.1}\\
\calA_{U_S}&=\frac{\Delta_{U,\uparrow}-\Delta_{U,\downarrow}}{2},\label{8.22.2}\\
2\delta(\rho_{S,i})\|A_S\|_{}&\geq 2L_{B}(\sigma_{S,i},U_{S}\rho_{S,i}U^{\dagger}_{S})\|A_S\|_{} \nt \\
&\geq \|\sigma_{S,i}-U_{S}\rho_{S,i}U^{\dagger}_{S}\|_{1}\|A_S\|_{} \nt\\
&\geq |\Delta_{i}-\Delta_{U,i}| ,\label{8.22.3}
\end{align}
where $i\in \{ \uparrow,\downarrow\}$, $\| X\|_1:=\Tr \sqrt{X^\dagger X}$ is the trace norm, and we used $\|\rho-\sigma\|_{1}\le2\sqrt{1-F^2(\rho,\sigma)}\le2L_{B}(\rho,\sigma)$ \cite{hayashi} in \eqref{8.22.3}.
Combining \eqref{8.22.1}--\eqref{8.22.3}, we obtain
\begin{align}
2\calA_{U_S}&=|\Delta_{U,\up}-\Delta_{U,\down}|\nonumber\\
&\le|\Delta_{\up}-\Delta_{\down}|+2(\delta (\rho_{S,\uparrow})+\delta (\rho_{S,\downarrow}))\|A_S\|_{}\nonumber\\
&=\Delta+2(\delta (\rho_{S,\uparrow})+\delta (\rho_{S,\downarrow}))\|A_S\|_{}.\label{12.19.1}
\end{align}

Hence, proving
\begin{align}
\delta (\rho_{S,\uparrow})+\delta (\rho_{S,\downarrow})\le2\delta (\rho_{S,\uparrow+\downarrow})\label{hukuhei}
\end{align}
suffices to show the desired inequality \eqref{help2}.
We firstly define some symbols.
We take purification $\psi_{SR_S,i}$ of $\rho_{S,i}$ $(i=\uparrow,\downarrow,\uparrow+\downarrow)$ such that $\rho_{R_S,\uparrow}:=\Tr_{S}[\psi_{SR_S,\uparrow}]$ and $\rho_{R_S,\downarrow}:=\Tr_{S}[\psi_{SR_S,\downarrow}]$ are pure states and orthogonal to each other.
In this case, $\ket{\psi_{SR_S,\uparrow+\downarrow}}=(\ket{\psi_{SR_S,\uparrow}}+\ket{\psi_{SR_S,\downarrow}})/\sqrt{2}$ holds.
We denote the purification of $\rho_E$ by $\psi_{ER_E}$.
We denote the initial and final states of the total system $SR_SER_E$ by $\psi_{tot,i}:=\psi_{SR_S,i}\otimes\psi_{ER_E}$ and $\psi'_{tot,i}:=U_{SE}\psi_{tot,i}U^{\dagger}_{SE}$, respectively.
We also denote the final states of $SR_S$ and $ER_E$ by $\sigma_{SR_S,i}:=\Tr_{ER_E}[\psi'_{tot,i}]$ and $\sigma_{ER_E,i}:=\Tr_{SR_S}[\psi'_{tot,i}]$, respectively.

We recall the fact that $\delta(\rho)$ is expressed in terms of fidelity:
\begin{align}
1-\frac{\delta (\rho_{S,i})^2}{2}=F(\sigma_{SR_S,i},U_S\psi_{SR_S,i}U^{\dagger}_S). \label{B12}
\end{align}
Uhlmann's theorem implies that there is a proper pure state $\phi'_{ER_E}$ on $ER_E$ such that
\begin{align}
&F(\sigma_{SR_S,\uparrow+\downarrow},U_S\psi_{SR_S,\uparrow+\downarrow}U^{\dagger}_S)\nt\\
&=|\bra{\psi_{tot,\uparrow+\downarrow}}U_S\ket{\psi_{SR_S,\uparrow+\downarrow}\otimes \phi'_{ER_E}}|.
\end{align}
Using $\ket{\psi_{SR_S,\uparrow+\downarrow}}=(\ket{\psi_{SR_S,\uparrow}}+\ket{\psi_{SR_S,\downarrow}})/\sqrt{2}$ and $\ket{\psi_{tot,\uparrow+\downarrow}}=(\ket{\psi_{tot,\uparrow}}+\ket{\psi_{tot,\downarrow}})/\sqrt{2}$, the right-hand side of the above relation is bounded from above as
\begin{align}
&|\bra{\psi_{tot,\uparrow+\downarrow}}U_S\ket{\psi_{SR_S,\uparrow+\downarrow}\otimes \phi'_{ER_E}}|\nt\\
&=\left|\frac{1}{2}\sum_{m=\uparrow,\downarrow}\bra{\psi_{tot,m}}U_S\ket{\psi_{SR_S,m}\otimes \phi'_{ER_E}}\right|\nonumber\\
&\le\frac{1}{2}\sum_{m=\uparrow,\downarrow}\left|\bra{\psi_{tot,m}}U_S\ket{\psi_{SR_S,m}\otimes \phi'_{ER_E}}\right|\nonumber\\
&\le\frac{1}{2}\sum_{m=\uparrow,\downarrow}F(\sigma_{SR_S,m},U_S\psi_{SR_S,m}U^{\dagger}_S)\label{B14}
\end{align}
Combining \eqref{B12}--\eqref{B14}, we obtain \eqref{hukuhei}.

\end{proofof}

\section{Proof of Lemma \ref{CL3}}\lb{s:pf-suf}

In this section, we prove Lemma \ref{CL3} in the main text.
In the proof, we use the following abbreviation for the convenience.
\begin{align}
\lambda_{\mathrm{diff}}(X)=\lambda_{\max}(X)-\lambda_{\min}(X).
\end{align}
where $\lambda_{\max}(X)$ and $\lambda_{\min}(X)$ is the maximum and minimum eigenvalues of $X$. Note that $\calA_{U_S}=\lambda_{\mathrm{diff}}(A_S-U^{\dagger}_SA_SU_S)/2$.

Lemma \ref{CL3} is easily given by the following two lemmas:
\begin{lemma}\label{CL3'}
We take an one-dimensional continuous system as ${\cal H}_E$, and set the position operator $x$ on it as the Hermitian $A_E$.
Given an arbitrary unitary $U_S:=\sum_{ij}u_{ij}\ket{i}\bra{j}$ on $S$ ($\ket{i}$ and $\ket{j}$ are the eigenvectors of $A_S$ whose eigenvalues are $h_i$ and $h_j$), and an arbitrary positive real number $\zeta$.
We define $U_{SE}$ and $\phi_{\zeta}$ as follows:
\begin{align}
\ket{\phi_{\zeta}}&:=C\int^{\infty}_{-\infty}\sqrt{e^{-\frac{x^{2}}{2\zeta^{2}}}}\ket{x}dx\\
U_{SE}&:=\sum_{ij}u_{ij}\ket{i}\bra{j}\otimes e^{-ip(h_i-h_j)}, \label{CL3-2'}
\end{align}
where $p$ represents the momentum operator and $C$ is a normalization constant.
By construction, $[U_{SE},A_S+A_E]=0$ is satisfied.

Then, for $\zeta\geq 9\calA_{U_S}/2\sqrt{2}$, the implementation set $\calI_{\zeta}=({\cal H_E}, A_E, \phi_{\zeta}, U_{SE})$ satisfies the following inequality for an arbitrary initial state $\rho_S$ on $S$:
\begin{align}
F_{e}(\rho_S,\Lambda_{U^{\dagger}_S}\circ\Lambda_S)&\ge \left|\left<T[e^{-(A'_S-A_S-h\hat{1})^2/8\zeta^{2}}]\right>_{\rho_S}\right|\label{CL3-3'}
\end{align}
where we set $\Lambda_{S}(\rho_S):=\Tr_{E}[U_{SE}(\rho_S\otimes\phi_{\zeta})U^{\dagger}_{SE}]$, $A'_S:=U^{\dagger}_SA_SU_S$, and $h$ is a real number satisfying $\|A'_S-A_S-h\hat{1}\|=\calA_{U_S}$.
The symbol $T$ represents the time-ordering product (e.g., $T[A'_SA_SA'_S]=A'^2_SA_S$).
\end{lemma}

\begin{lemma}\label{C2}
We consider a quantum system, and take an arbitrary Hermitian $X$ and arbitrary unitary $U$ on the system.
For $X$ and $X':=U^{\dagger}XU$, we define $X_0:=x_0\hat{1}$ such that $\|X-X'-X_0\|=\lambda_{\mathrm{diff}}(X-X')/2$.
When $\|X-X'-X_0\|\le\|X\|\le a$ holds for a positive number $a\le1/9$, 
the following inequality holds:
\begin{align}
\min_{\rho}\left|\left<T[e^{-(X-X'-X_0)^2}]\right>_{\rho}\right|&\ge 1-\frac{\lambda_{\mathrm{diff}}(X-X')^{2}}{4}(1+2a)^2\label{C2-1}.
\end{align}
\end{lemma}

These two lemmas directly implies Lemma \ref{CL3} by subsituting $A_S/(2\sqrt{2}\zeta)$ and $U_S$ for $X$ and $U$, respectively.

\begin{proofof}{Lemma \ref{CL3'}}
We expand the initial state as
\begin{align}
\rho_S&:=\sum_{\lambda}p_{\lambda}\ket{\psi^{(\lambda)}_S}\bra{\psi^{(\lambda)}_S},\\
\ket{\psi^{(\lambda)}_S}&:=\sum_{i}a^{(\lambda)}_{i}\ket{i},\\
\ket{\psi_{SR}}&:=\sum_{i}\sqrt{p^{(\lambda)}}a^{(\lambda)}_{i}\ket{i}\ket{\lambda}_R.
\end{align}
Using these symbols, we can describe $F_{e}(\rho_S,\Lambda_{U^{\dagger}_S}\circ\Lambda_S)$ as follows:
\begin{align}
&F_{e}(\rho_S,\Lambda_{U^{\dagger}_S}\circ\Lambda_S)^2\nt\\
&=\bra{\psi_{SR}}\Tr_{E}[U^{\dagger}_SU_{SE}(\psi_{SR}\otimes\phi_{\zeta})U^{\dagger}_{SE}U_S]\ket{\psi_{SR}}
\end{align}
By defining shifted $\phi_{\zeta}$ by $h$ as
\begin{align}
\ket{\phi_{\zeta,h}}:=e^{-iph}\ket{\phi_{\zeta}}
\end{align}
with the momentum operator $p$ in ${\cal H}_E$, we construct an orthonomal basis of $E$ as $\{\ket{\phi_{\zeta,h}},\ket{\phi^{(1)}_{E}},...\}$.
Expanding $\ket{\phi_E}$ with this basis, we find
\begin{align}
F_{e}&(\rho_S,\Lambda_{U^{\dagger}_S}\circ\Lambda_S)^2\nonumber\\
=&\bra{\psi_{SR}}\Tr_{E}[U^{\dagger}_SU_{SE}(\psi_{SR}\otimes\phi_{\zeta})U^{\dagger}_{SE}U_S]\ket{\psi_{SR}}\nonumber\\
=&\bra{\psi_{SR}}\bra{\phi_{\zeta,h}}U^{\dagger}_SU_{SE}(\psi_{SR}\otimes\phi_{\zeta})U^{\dagger}_{SE}U_S\ket{\psi_{SR}}\ket{\phi_{\zeta,h}}\nt\\
&+\bra{\psi_{SR}}\bra{\phi^{(1)}_{E}}U^{\dagger}_SU_{SE}(\psi_{SR}\otimes\phi_{\zeta})U^{\dagger}_{SE}U_S\ket{\psi_{SR}}\ket{\phi^{(1)}_E}\nt\\
&+...\nonumber\\
\ge&\bra{\psi_{SR}}\bra{\phi_{\zeta,h}}U^{\dagger}_SU_{SE}(\psi_{SR}\otimes\phi_{\zeta})U^{\dagger}_{SE}U_S\ket{\psi_{SR}}\ket{\phi_{\zeta,h}} \nt \\
\ge& |\bra{\psi_{SR}}\bra{\phi_{\zeta,h}}U^{\dagger}_SU_{SE}\ket{\psi_{SR}\otimes\phi_{\zeta}}|^2,
\end{align}
which is equivalent to
\begin{align}
F_{e}(\rho_S,\Lambda_{U^{\dagger}_S}\circ\Lambda_S)\ge |\bra{\phi_{\zeta,h}}\bra{\psi_{SR}}U^{\dagger}_SU_{SE}\ket{\psi_{SR}}\ket{\phi_{\zeta}}|. \label{B7}
\end{align}

Using \eqref{CL3-2'} and the following equality:
\begin{align}
\braket{\phi_{\zeta,\Delta}|\phi_{\zeta}}&=C^2\int^{\infty}_{-\infty}e^{-\frac{(x-\Delta)^2}{4\zeta^{2}}}e^{-\frac{x^2}{4\zeta^{2}}}dx\nonumber\\
&=C^2\int^{\infty}_{-\infty}e^{-\frac{(x-\frac{\Delta}{2})^2+\frac{\Delta^2}{4}}{2\zeta^{2}}}dx\nonumber\\
&=e^{-\frac{\Delta^2}{8\zeta^{2}}}, \label{B8}
\end{align}
we can transform the right hand side of \eqref{B7} as
\begin{align}
&|\bra{\phi_{\zeta,h}}\bra{\psi_{SR}}U^{\dagger}_SU_{SE}\ket{\psi_{SR}}\ket{\phi_{\zeta}}|\nt\\
&=\left|\sum_{\lambda}p_{\lambda}\sum_{ijk}a^{(\lambda)}_{i}u_{ji}a^{(\lambda)*}_{k}u^{*}_{jk}e^{-\frac{(h_i-h_j-h)^2}{8\zeta^{2}}}\right|\nonumber\\
&=\left|\sum_{\lambda}p_{\lambda}\sum_{ijk}a^{(\lambda)}_{i}u_{ji}a^{(\lambda)*}_{k}u^{*}_{jk}\sum^{\infty}_{l=0}\frac{(-1)^l}{l!}\left(\frac{(h_i-h_j-h)^2}{8\zeta^{2}}\right)^l\right|.\label{CLp5}
\end{align}
Finally, applying
\begin{align}
&\sum_{ijk}a^{(\lambda)}_{i}u_{ji}a^{(\lambda)*}_{k}u^{*}_{jk}\left(\frac{(h_i-h_j-h)^2}{8\zeta^{2}}\right)^l\nt\\
&=\bra{\psi^{(\lambda)}_{S}}T\left[\left(\frac{(A'_S-A_S-h\hat{I})^2}{8\zeta^{2}}\right)^l\right] \ket{\psi^{(\lambda)}_{S}}\label{B10}
\end{align}
to the right-hand side of \eqref{CLp5}, we obtain
\begin{align}
&\left|\sum_{\lambda}p_{\lambda}\sum_{ijk}a^{(\lambda)}_{i}u_{ji}a^{(\lambda)*}_{k}u^{*}_{jk}\sum^{\infty}_{l=0}\frac{(-1)^l}{l!}\left(\frac{(h_i-h_j-h)^2}{8\zeta^{2}}\right)^l\right|\nt\\
&=\left|\sum_{\lambda}p_{\lambda}\bra{\psi^{(\lambda)}_S}\sum^{\infty}_{l=0}\frac{(-1)^l}{l!}T\left[\left(\frac{(A'_S-A_S-h\hat{I})^2}{8\zeta^{2}}\right)^l\right]\ket{\psi^{(\lambda)}_S}\right|\nonumber\\
&=\left|\left<T[e^{-(A'_S-A_S-h\hat{I})^2/8\zeta^{2}}]\right>_{\rho}\right|.
\end{align}
In conclusion, we arrive at the desired inequality
\begin{align}
F_{e}(\rho_S,\Lambda_{U^{\dagger}_S}\circ\Lambda_S)\ge \left|\left<T[e^{-(A'_S-A_S-h\hat{I})^2/8\zeta^{2}}]\right>_{\rho}\right|.
\end{align}
\end{proofof}

\begin{proofof}{Lemma \ref{C2}}
The Taylor expansion of $T[e^{-(X-X'-X_0)^2}]$ reads
\begin{align}
T[e^{-(X-X'-X_0)^2}]=1-T_2+\frac{1}{2!}T_4-\frac{1}{3!}T_6+...,
\end{align}
where we employed the abbreviation $T_m:=T[(X-X'-X_0)^m]$.
Using this expression, we have the following inequality:
\begin{align}
\left|\left<T[e^{-(X-X'-X_0)^2}]\right>_{\rho}\right|&\ge
\left|\mbox{Re}\left(\left<T[e^{-(X-X'-X_0)^2}]\right>_{\rho}\right)\right|\nonumber\\
&\ge1-\sum^{\infty}_{m=1}\frac{1}{m!}\left|\left<S_{2m}\right>_{\rho}\right|\nonumber\\
&\ge1-\sum^{\infty}_{m=1}\frac{1}{m!}\|S_{2m}\|,
\end{align}
where we denoted the Hermitian part of $T_{2m}$ by $S_{2m}:=(T_{2m}+T^{\dagger}_{2m})/{2}$.
For convenience, we also define the anti-Hermitian part of $T_{sm}$ as $A_{2m}$.
The operator norms of $S_{2m}$ and $A_{2m}$ are bounded from above as:
\begin{align}
\|S_{2m}\|&\le\|X-X'-X_0\|^2(6a)^{m-1}\label{pC2A}\\
\|A_{2m}\|&\le\|X-X'-X_0\|(6a)^{m-1}\label{pC2B}.
\end{align}
We first see how \eqref{pC2A} leads to the desired inequality \eqref{C2-1}, and then we prove the above inequalities.
Using \eqref{pC2A}, the left-hand side of \eqref{C2-1} is evaluated as
\begin{align}
&\min_{\rho}\left|\left<T[e^{-(X-X'-X_0)^2}]\right>_{\rho}\right|\nonumber\\
&\ge1-\|X-X'-X_0\|^2\left(1+\frac{1}{2!}(6a)+\frac{1}{3!}(6a)^2+...\right)\nonumber\\
&=1-\|X-X'-X_0\|^2\frac{e^{6a}-1}{6a}\nonumber\\
&=1-\frac{\lambda_{\mathrm{diff}}(X-X')^2}{4}\frac{e^{6a}-1}{6a}\nonumber\\
&\ge1-\frac{\lambda_{\mathrm{diff}}(X-X')^2}{4}(1+2a)^2,
\end{align}
where we used $a\le1/9$ in the last inequality.

We now prove \eqref{pC2A} by using the mathematical induction on $m$.
We also prove \eqref{pC2B} as a by-product.
In this proof, we put $Y:=X-X'-X_0$ for brevity.
We first show \eqref{pC2A} and \eqref{pC2B} for $m=1$.
Recalling $T_2=T[(X-X'-X_0)^2]$, we have
\begin{align}
S_2=Y^2,\enskip A_2=[X',X]=[Y,X],
\end{align}
which directly imply
\begin{align}
\|S_2\|=\|Y\|^2,\enskip \|A_2\|\le 2a\|Y\|\le\|Y\|.
\end{align}
Hence, \eqref{pC2A} and \eqref{pC2B} hold for $m=1$.

We next show the inductive step.
Assume that \eqref{pC2A} and \eqref{pC2B} hold for $m\le k$.
We shall show that \eqref{pC2A} and \eqref{pC2B} also hold for $m=k+1$.
Inserting the following recursion twice to the definition of $T_{2m}$
\begin{align}
T_{n}=T_{n-1}X-(X'+X_0)T_{n-1}=[T_{n-1},X]+YT_{n-1},
\end{align}
we obtain the recursion relation between $T_{2n}$ and $T_{2(n-1)}$:
\begin{align}
T_{2n}=&[T_{2n-1},X]+YT_{2n-1}\nonumber\\
=&[[T_{2(n-1)},X],X]+[YT_{2(n-1)},X]\nonumber\\
&+Y[T_{2(n-1)},X]+Y^2T_{2(n-1)}.
\end{align}
Using the relations $T_{2(n-1)}=S_{2(n-1)}+A_{2(n-1)}$, $S_{2n}=\frac{T_{2n}+T^{\dagger}_{2n}}{2}$ and $A_{2n}=\frac{T_{2n}-T^{\dagger}_{2n}}{2}$, we divide this recursion into the ones about $S_{2n}$ and $A_{2n}$, respectively.
From $T_{2(n-1)}=S_{2(n-1)}+A_{2(n-1)}$, we obtain
\begin{align}
T_{2n}=&[[S_{2(n-1)},X],X]+[YS_{2(n-1)},X]\nonumber\\
&+Y[S_{2(n-1)},X]+Y^2S_{2(n-1)}\nonumber\\
&+[[A_{2(n-1)},X],X]+[YA_{2(n-1)},X]\nonumber\\
&+Y[A_{2(n-1)},X]+Y^2A_{2(n-1)}\nonumber\\
=&[[S_{2(n-1)},X],X]+2YS_{2(n-1)}X\nonumber\\
&-XYS_{2(n-1)}-YXS_{2(n-1)}+Y^2S_{2(n-1)}\nonumber\\
&+[[A_{2(n-1)},X],X]+2YA_{2(n-1)}X\nonumber\\
&-XYA_{2(n-1)}-YXA_{2(n-1)}+Y^2A_{2(n-1)}.
\end{align}
From $S_{2n}=\frac{T_{2n}+T^{\dagger}_{2n}}{2}$ and $A_{2n}=\frac{T_{2n}-T^{\dagger}_{2n}}{2}$, above equality reads
\begin{align}
S_{2n}=&[[S_{2(n-1)},X],X]\nonumber\\
&+\frac{1}{2}(2YS_{2(n-1)}X-\{X,Y\}S_{2(n-1)}+Y^2S_{2(n-1)})\nonumber\\
&+\frac{1}{2}(2XS_{2(n-1)}Y-S_{2(n-1)}\{X,Y\}+S_{2(n-1)}Y^2)\nonumber\\
&+\frac{1}{2}(2YA_{2(n-1)}X-\{X,Y\}A_{2(n-1)}+Y^2A_{2(n-1)})\nonumber\\
&-\frac{1}{2}(2XA_{2(n-1)}Y-A_{2(n-1)}\{X,Y\}+A_{2(n-1)}Y^2)\nonumber\\
A_{2n}=&[[A_{2(n-1)},X],X]\nonumber\\
&+\frac{1}{2}(2YS_{2(n-1)}X-\{X,Y\}S_{2(n-1)}+Y^2S_{2(n-1)})\nonumber\\
&-\frac{1}{2}(2XS_{2(n-1)}Y-S_{2(n-1)}\{X,Y\}+S_{2(n-1)}Y^2)\nonumber\\
&+\frac{1}{2}(2YA_{2(n-1)}X-\{X,Y\}A_{2(n-1)}+Y^2A_{2(n-1)})\nonumber\\
&+\frac{1}{2}(2XA_{2(n-1)}Y-A_{2(n-1)}\{X,Y\}+A_{2(n-1)}Y^2).
\end{align}
By taking operator norms of above relations, the following recursion relations are obtained:
\begin{align}
\|S_{2n}\|\le&\|S_{2(n-1)}\|4\|X\|^2+\|S_{2(n-1)}\|(4\|X\|\|Y\|+\|Y\|^2)|\nonumber\\
&+\|A_{2(n-1)}\|(4\|X\|\|Y\|+\|Y\|^2)|\nonumber\\
\le&\|S_{2(n-1)}\|9a^2+\|A_{2(n-1)}\|5a\|Y\|,\\
\|A_{2n}\|\le&\|A_{2(n-1)}\|4\|X\|^2+\|S_{2(n-1)}\|(4\|X\|\|Y\|+\|Y\|^2)|\nonumber\\
&+\|A_{2(n-1)}\|(4\|X\|\|Y\|+\|Y\|^2)|\nonumber\\
&\le\|A_{2(n-1)}\|9a^2+\|S_{2(n-1)}\|5a\|Y\| .
\end{align}
Finally, using the induction hypothesis, we find that \eqref{pC2A} and \eqref{pC2B} hold for $m=k+1$:
\begin{align}
\|S_{2(k+1)}\|&\le\|Y\|^2(6a)^{k-1}9a^2+\|Y\|(6a)^{k-1}5a\|Y\|\nonumber\\
&\le\|Y\|^2(6a)^{k}\left(\frac{9a}{6}+\frac{5}{6}\right)\nonumber\\
&\le\|Y\|^2(6a)^{k},\\
\|A_{2(k+1)}\|&\le\|Y\|(6a)^{k-1}9a^2+\|Y\|^2(6a)^{k-1}5a\|Y\|\nonumber\\
&\le\|Y\|(6a)^{k}\left(\frac{9a}{6}+\frac{a}{6}\right)\nonumber\\
&\le\|Y\|(6a)^{k}.
\end{align}
By mathematical induction, \eqref{pC2A} and \eqref{pC2B} hold for any $m$.
\end{proofof}

\section{Proof of \eqref{true}}\lb{s:pf-true}

\begin{proofof}{\eqref{true}}
We firstly consider the case where both of $\rho_S$ and $\rho_E$ are pure states.
We here reshow some definitions of symbols used in this proof.
We denote the desired final state and realized final state by
\begin{align}
\ket{\rho'_S}&:=U_S\ket{\rho_S},\\
\ket{\Psi'_{SE}}&:=U_{SE}\ket{\rho_{S}}\ket{\rho_E},\\
\sigma_{E}&:=\Tr_{S}[\Psi'_{SE}], \\
\sigma_{S}&:=\Tr_{E}[\Psi'_{SE}].
\end{align}
In a similar manner, the final state of $\psi_i$ is denoted by
\begin{align}
\ket{\psi'_{i}}&:=U_S\ket{\psi_{i}}\\
\ket{\Psi'_{i,SE}}&:=U_{SE}\ket{\psi_{i}}\ket{\rho_E},\\
\sigma_{i,E}&:=\Tr_{S}[\Psi'_{i,SE}], \\
\sigma_{i,S}&:=\Tr_{E}[\Psi'_{i,SE}].
\end{align}
With noting the definition of $\delta (\rho)$, Uhlmann's theorem tells that there are two pure states $\phi'_{E}$ and $\phi'_{i,E}$ satisfying
\begin{align}
|\braket{\Psi'_{SE}|\rho'_S\otimes \phi'_{E}}|&=(1-\frac{\delta (\rho_S)^2}{2})\\
|\braket{\Psi'_{i,SE}|\psi'_{i}\otimes \phi'_{i,E}}|&=(1-\frac{\delta (\psi_{i})^2}{2}),
\end{align}
where we wrote $\ket{\rho}\otimes \ket{\phi}$ as $\ket{\rho \otimes \phi}$.

In the case of a pure state, $2V_{A_E}(\rho_E)=\sqrt{\calF(\rho_E)}$ holds by definition, and the pure state $\rho_S$ is written as $\ket{\rho_S}=\sum_{i}\alpha_{i}\ket{\psi_{i}}$.
In this case, $r_i=|\alpha_i|^2$ holds, and the desired inequality \eqref{true} follows from
%\begin{align}V_{A_E}(\rho_E)\ge\frac{\chi(\rho_S, \{ \psi_i\} )}{6\overline{\delta}(\rho_S, \{ \psi_i\} )}-3\|A_S\| , \label{true'}\end{align}
\begin{align}
\chi( \rho_S, \{ \psi_i\} )^2 \le 64 \overline{\delta}(\rho_S, \{ \psi_i\} )^2(V_{A_E}(\rho_E)+2\|A_S\|)^2, \label{5}
\end{align}
where we used the abbreviation:
\balign{
\overline{\delta}^2(\rho_S, \{ \psi_i\} )&:=\delta_{}(\rho_S)^2+\sum_ir_i\delta_{}(\psi_i)^2.
}
Since the above inequality \eqref{5} reduces to a trivial relation $V_{A_E}(V_E)\ge0$ when $\overline{\delta }(\rho_S, \{ \psi_i\} )>\chi( \rho_S, \{ \psi_i\} )^2/{20\|A_S\|}$ is satisfied, in the following we prove \eqref{5} only for the case of $\overline{\delta}(\rho_S, \{ \psi_i\} )\le \chi( \rho_S, \{ \psi_i\} )^2/{20\|A_S\|}$.

We first employ \eqref{c-single} (the relationship (c)), which is repeated below:
\balign{
\chi(\rho_S, \{ \psi_i\})\le\sqrt{\sum_{i} r_i \Delta_i ^2}+\Delta'+4\overline{\delta}(\rho_S)\|A_S\|,\label{D13}
}
where $\Delta_{i}:=|\left<A_E\right>_{\sigma_{i,E}}-\left<A_E\right>_{\phi'_E}|$ and $\Delta':=|\left<A_E\right>_{\sigma_{E}}-\left<A_E\right>_{\phi'_E}|$.
The equation \eqref{c-single} is shown as follows:
\begin{widetext}
\begin{align}
\chi(\rho_S, \{ \psi_i\})^2=&\sum_{i} r_i(\left<A_S\right>_{\psi_{i}}-\left<A_S\right>_{U_S\psi_{i}U^{\dagger}_S}+\left<A_S\right>_{U_S\rho_SU^{\dagger}_S} -\left<A_S\right>_{\rho_S})^2\nonumber\\
\le&
\sum_{i} r_i\big(|\left<A_S\right>_{\psi_{i}}-\left<A_S\right>_{\sigma_{i,S}}+\left<A_S\right>_{\sigma_{S}}-\left<A_S\right>_{\rho_S}|+2(\delta (\psi_{i})+\delta (\rho_S))\|A_S\|\big)^2,\nonumber\\
=&\sum_{i} r_i\big(|\left<A_E\right>_{\sigma_{i,E}}-\left<A_E\right>_{\rho_E}+\left<A_E\right>_{\rho_E}-\left<A_E\right>_{\sigma_E}|+2(\delta (\psi_{i})+\delta (\rho_S))\|A_S\|\big)^2\nonumber\\
=&\sum_{i} r_i(|\left<A_E\right>_{\sigma_{i,E}}-\left<A_E\right>_{\sigma_E}|+2(\delta (\psi_{i})+\delta (\rho_S))\|A_S\|)^2\nonumber\\
\le&\left(\sqrt{\sum_{i} r_i(\left<A_E\right>_{\sigma_{i,E}}-\left<A_E\right>_{\sigma_E})^2}+2\sqrt{\sum_{i} r_i(\delta (\psi_{i})+\delta (\rho_S))^2\|A_S\|^2}\right)^2 \nt \\
\le&\left(\sqrt{\sum_{i} r_i(\left<A_E\right>_{\sigma_{i,E}}-\left<A_E\right>_{\phi'_E})^2}+|\left<A_E\right>_{\sigma_{E}}-\left<A_E\right>_{\phi'_E}|+2\sqrt{\sum_{i} r_i(\delta (\psi_{i})+\delta (\rho_S))^2\|A_S\|^2}\right)^2\nonumber\\
=&\left(\sqrt{\sum_{i} r_i \Delta_i ^2}+\Delta'+2\sqrt{\sum_{i} r_i(\delta (\psi_{i})+\delta (\rho_S))^2\|A_S\|^2}\right)^2
\label{18},
\end{align}
\end{widetext}
where we used the conservation of $A$ in the total system in the third line.
We obtain \eqref{D13} since the third term in the right-hand side of \eqref{18} is easily bounded as
\begin{align}
\sum_{i} r_i(\delta (\psi_{i})+\delta (\rho_S))^2\|A_S\|^2
\le4\overline{\delta}(\rho_S, \{ \psi_i\} )^2\|A_S\|^2.
\end{align}

Below we consider the first term in the right-hand side of \eqref{18}.
This term is evaluated by using the Lemma.~\ref{L3}, or \eqref{L3-single}, 
\begin{align}
&\sum_i r_i \Delta_i^2=\sum_{i} r_i(\left<A_E\right>_{\sigma_{i,E}}-\left<A_E\right>_{\phi'_E})^2\nonumber\\
&\le\sum_{i} r_iL(\sigma_{i,E},\phi'_E)^2(V_{A_E}(\sigma_{i,E})+V_{A_E}(\phi'_{E})+\Delta_i)^2 \label{19}
\end{align}
The first term in the bracket, $V_{A_E}(\sigma_{i,E})$, is bounded as \eqref{b-single} (relationship (b)):
\eq{
V_{A_E}(\sigma_{i,E})\leq V_{A_E}(\rho_E)+\|A_S\| .
}
In the subsequent analysis, we first bound the two correction terms, $V_{A_E}(\phi'_{E})$ and $|\left<A_E\right>_{\sigma_{i,E}}-\left<A_E\right>_{\phi'_E}|$, by quantities independent of $i$.
We then evaluate $\sum_{i} r_iL(\sigma_{i,E},\phi'_E)^2$ by using \eqref{a-single} (relationship (a)).

We first derive the bound for $V_{A_E}(\phi'_{E})$.
We compare the fluctuation of $A$ in $\rho_{S}\otimes\rho_E$ and $\rho'_{S}\otimes\phi'_{E}$:
\begin{align}
&V_{A_S+A_E}(\rho_{S}\otimes\rho_E)^2 \nt \\
=&V_{A_S+A_E}(\Psi'_{SE})^2 \nt \\
=&\Tr[(A_S+A_E-\left<A_S+A_E\right>_{\Psi'_{SE}})^2\Psi'_{SE}]\nonumber\\
\ge&\Tr[P_{\rho'_{S}\otimes\phi'_{E}}(A_S+A_E-\left<A_S+A_E\right>_{\Psi'_{SE}})^2P_{\rho'_{S}\otimes\phi'_{E}}\Psi'_{SE}]\nonumber\\
\ge&\Tr[(A_S+A_E-\left<A_S+A_E\right>_{\Psi'_{SE}})^2P_{\rho'_{S}\otimes\phi'_{E}}\Psi'_{SE}P_{\rho'_{S}\otimes\phi'_{E}}]\nonumber\\
=&\Tr[(A_S+A_E-\left<A_S+A_E\right>_{\Psi'_{SE}})^2\rho'_{S}\otimes\phi'_{E}]\Tr[P_{\rho'_{S}\otimes\phi'_{E}}\Psi'_{SE}]\nonumber\\
=&\Tr[(A_S+A_E-\left<A_S+A_E\right>_{\Psi'_{SE}})^2\rho'_{S}\otimes\phi'_{E}]\left(1-\frac{\delta (\rho_S)^2}{2}\right)^2\nonumber\\
\ge&V_{A_S+A_E}(\rho'_{S}\otimes\phi'_{E})^2 \left(1-\frac{\delta (\rho_S)^2}{2}\right)^2.
\end{align}
Here, $P_{\rho'_{S}\otimes\phi'_{E}}:=\ket{\rho'_{S}\otimes\phi'_{E}}\bra{\rho'_{S}\otimes\phi'_{E}}$ is the projection operator onto ${\rho'_{S}\otimes\phi'_{E}}$.
By substituting $V_{A_S+A_E}(\rho_{S}\otimes\rho_E)^2=V_{A_S}(\rho_S)^2+V_{A_E}(\rho_E)^2 \leq \|A_S\| ^2/4 +V_{A_E}(\rho_E)^2$ and $V_{A_S+A_E}(\rho'_{S}\otimes\phi'_{E})^2= V_{A_S}(\rho'_S)^2+ V_{A_E}(\phi'_E)^2 \geq V_{A_E}(\phi'_E)^2 $, we arrive at an upper bound for $V_{A_E}(\phi'_E)$:
\begin{align}
V_{A_E}(\phi'_E)\le\frac{\sqrt{\frac{1}{4}\|A_S\|^2+V_{A_E}(\rho_E)^2}}{1-\frac{\delta (\rho_S)^2}{2}}.
\end{align}

We next derive the bound for $\Delta_i:= |\left<A_E\right>_{\sigma_{i,E}}-\left<A_E\right>_{\phi'_E}|$.
With noting our condition $\overline{\delta}(\rho_S, \{ \psi_i\} )\le\frac{V_{A'_S-A_S}(\rho_S)}{20\|A_S\|}\le1/20\le1$, we evaluate $|\left<A_E\right>_{\sigma_{i,E}}-\left<A_E\right>_{\phi'_E}|$ as
\begin{align}
&|\left<A_E\right>_{\sigma_{i,E}}-\left<A_E\right>_{\phi'_E}|\nonumber\\
\le&|\left<A_E\right>_{\sigma_{i,E}}-\left<A_E\right>_{\sigma_E}|+|\left<A_E\right>_{\sigma_E}-\left<A_E\right>_{\phi'_E}|\nonumber\\
=&|\left<A_S\right>_{\psi_{i}}-\left<A_S\right>_{\sigma_{i,S}}-\left<A_S\right>_{\sigma_S}+\left<A_S\right>_{\rho_S}|+\Delta'\nonumber\\
\le&2\|A_S\|+\Delta'.
\end{align}
In the third line, we used the fact that the minimal eigenvalue of $A_S$ is zero.

From Lemma \ref{L3} and $L(\sigma_{E},\phi'_{E})\le\delta(\rho_S)$, we derive the bound for $\Delta'$ as follows
\begin{align}
\Delta'\le&\frac{L(\sigma_{E},\phi'_{E})}{1-L(\sigma_{E},\phi'_{E})}(V_{A_E}(\sigma_E)+V_{A_E}(\phi'_E))\nonumber\\
\le&\frac{\delta(\rho_S)}{1-\delta(\rho_S)}\nonumber\\
&\times\left(V_{A_E}(\rho_E)+\|A_S\|+\frac{\sqrt{\frac{1}{4}\|A_S\|^2+V_{A_E}(\rho_E)^2}}{1-\frac{\delta (\rho_S)^2}{2}}\right)
\end{align}

At present, we have an upper bound for the right-hand side of \eqref{19} as
\balign{
&\sum_{i} r_iL(\sigma_{i,E},\phi'_E)^2(V_{A_E}(\sigma_{i,E})+V_{A_E}(\phi'_{E})+\Delta_i)^2 \nt \\
\leq& \left[ \frac{1}{1-\delta (\rho_S)}V_{A_E}(\rho_E)+\frac{3-2\delta (\rho_S)}{1-\delta (\rho_S)}\|A_S\| \right. \nonumber\\
&\left.+ \frac{1}{1-\delta (\rho_S)}\frac{\sqrt{\frac{1}{4}\|A_S\|^2+V_{A_E}(\rho_E)^2}}{1-\frac{\delta (\rho_S)^2}{2}} \right] ^2 \sum_{i} r_iL(\sigma_{i,E},\phi'_E)^2
}
We finally calculate the bound for $\sum_{i} r_iL(\sigma_{i,E},\phi'_E)^2$.
Using a relation $\Tr[AB]\geq \Tr[\rho A\rho B]$ for positive Hermite operators $A$, $B$ and a density matrix $\rho$ repeatedly, we have
\begin{align}
&r_i\bra{\phi'_{E}}\sigma_{i,E}\ket{\phi'_{E}}\nonumber\\
=&\Tr[(\rho'_S\otimes\phi'_E)(\psi'_{i}\otimes\sigma_{i,E})]\nonumber\\
\ge&\Tr[\Psi'_{SE}(\rho'_S\otimes\phi'_E)\Psi'_{SE}(\psi'_{i}\otimes\sigma_{i,E})]\nonumber\\
=&\Tr[\Psi'_{SE}(\rho'_S\otimes\phi'_E)]\Tr[\Psi'_{SE}(\psi'_{i}\otimes\sigma_{i,E})]\nonumber\\
\ge&\Tr[\Psi'_{SE}(\rho'_S\otimes\phi'_E)]\Tr[\Psi'_{SE}\Psi'_{i,SE}(\psi'_{i}\otimes\sigma_{i,E})\Psi'_{i,SE}]\nonumber\\
=&\Tr[\Psi'_{SE}(\rho'_S\otimes\phi'_E)]\Tr[\Psi'_{SE}\Psi'_{i,SE}]\Tr[(\psi'_{i}\otimes\sigma_{i,E})\Psi'_{i,SE}]\nonumber\\
\ge&\Tr[\Psi'_{SE}(\rho'_S\otimes\phi'_E)]\Tr[\Psi'_{SE}\Psi'_{i,SE}]\nonumber\\
&\times\Tr[(\psi'_{i}\otimes\phi'_{i,E})(\psi'_{i}\otimes\sigma_{i,E})(\psi'_{i}\otimes\phi'_{i,E})\Psi'_{i,SE}]\nonumber\\
=&\Tr[\Psi'_{SE}(\rho'_S\otimes\phi'_E)]\Tr[\Psi'_{SE}\Psi'_{i,SE}]\nonumber\\
&\times\Tr[(\psi'_{i}\otimes\phi'_{i,E})(\psi'_{i}\otimes\sigma_{i,E})]\Tr[(\psi'_{i}\otimes\phi'_{i,E})\Psi'_{i,SE}]\nonumber\\
=&\Tr[\Psi'_{SE}(\rho'_S\otimes\phi'_E)]\Tr[\Psi'_{SE}\Psi'_{i,SE}]\nonumber\\
&\times\Tr[(\psi'_{i}\otimes\phi'_{i,E})\Psi'_{i,SE}]\Tr[\phi'_{i,E}\sigma_{i,E}]\nonumber\\
\ge& r_i \left(1-\frac{\delta (\rho_S)^2}{2}\right)^2\left(1-\frac{\delta (\psi_{i})^2}{2}\right)^4.
\end{align}
We combine $\bra{\phi'_{E}}\sigma_{i,E}\ket{\phi'_{E}}=(1-{L(\phi'_{E},\sigma_{i,E})^2}/{2})^2$ and 
\begin{align}
&\sum_{i} r_i \left(1-\frac{\delta (\rho_S)^2}{2}\right)^2\left(1-\frac{\delta (\psi_{i})^2}{2}\right)^4\nonumber\\
&\ge\left(1-\frac{\delta (\rho_S)^2}{2}\right)^2\left(1-\frac{\sum_{i} r_i \delta (\psi_{i})^2}{2}\right)^4,
\end{align}
which follows from Jensen's inequality, with noting $0\le1-L(\phi'_{E},\sigma_{i,E})^2/2\le1$, and obtain
\begin{align}
1-\frac{\sum_{i} r_i L(\phi'_{E},\sigma_{i,E})^2}{2}\ge1-(\delta (\rho_S)^2+2\sum_{i} r_i \delta (\psi_{i})^2),
\end{align}
which directly implies the desired bound
\begin{align}
\sum_{i} r_i L(\phi'_{E},\sigma_{i,E})^2&\le2(\delta (\rho_S)^2+2\sum_{i} r_i \delta (\psi_{i})^2)\nonumber\\
&\le 4\overline{\delta}(\rho_S, \{ \psi_i\} )^2.
\end{align}

In summary, by substituting all the obtained results into the right-hand side of \eqref{18} and with noting $\delta(\rho_S)\le\overline{\delta}(\rho_S,\{\psi_i\})$, we obtain the following inequality
\begin{align}
\chi(\rho_S, \{ \psi_i\} ) &\le 2\overline{\delta}(\rho_S, \{ \psi_i\} ) \nonumber\\
&\times\left[ \frac{1.5}{1-\delta (\rho_S)}V_{A_E}(\rho_E)+\frac{5.5-4\delta (\rho_S)}{1-\delta (\rho_S)}\|A_S\| \right.
\nonumber\\
&\left.+ \frac{1.5}{1-\delta (\rho_S)}\frac{\sqrt{\frac{1}{4}\|A_S\|^2+V_{A_E}(\rho_E)^2}}{1-\frac{\delta (\rho_S)^2}{2}} \right]
\end{align}
We note the inequality $\delta (\rho_S)\le1/20$, which follows from the condition $\overline{\delta}(\rho_S, \{ \psi_i\} )\le {\chi(\rho_S, \{ \psi_i\} )}/{20\|A_S\|}\leq 1/20$.
Then, using the relation $\sqrt{\frac{1}{4}\|A_S\|^2+V_{A_E}(\rho_E)^2}\le\frac{\|A_S\|}{2}+V_{A_E}(\rho_E)$, we obtain:
\begin{align}
&\chi(\rho_S, \{ \psi_i\} )\nonumber\\
&\le 2\overline{\delta}(\rho_S, \{ \psi_i\} ) (3.159...\times V_{A_E}(\rho_E)+6.369...\times \|A_S\|)\nonumber\\
&\le \overline{\delta}(\rho_S, \{ \psi_i\} ) (5\sqrt{2}V_{A_E}(\rho_E)+10\sqrt{2}\|A_S\|), \label{D23}
\end{align}
which readily implies \eqref{5} for the case where both of $\rho_S$ and $\rho_E$ are pure.

%%%%%%%%%%%%%%%%%%%

\bigskip

Next, we consider the case where $\rho_S$ is mixed and $\rho_E$ is pure.
We take a purification of $\rho_S$ as $\ket{\psi_{SR}}:=\sum_{\lambda,i}\sqrt{p_{\lambda}}\ket{\lambda}\ket{\rho_\lambda}$.
We expand each pure state $\rho_\lambda$ with the orthogonal basis $\{\psi_{i}\}$ as $\ket{\rho_\lambda}=\sum_{i}\alpha^{(\lambda)}_{i}\ket{\psi_{i}}$.
Then, $\ket{\psi_{SR}}$ is rewritten as 
\begin{align}
\ket{\psi_{SR}}=\sum_{i,\lambda}\sqrt{p_\lambda}\alpha^{(\lambda)}_i\ket{\lambda}\ket{\psi_{i}}.
\end{align}
By setting $\{\sqrt{p_\lambda}\alpha^{(\lambda)}_i\}$ and $\{\ket{\lambda}\ket{\psi_{i}}\}$ to $\{\alpha_i\}$ and $\{\ket{\psi_{i}}\}$ in the derivation of \eqref{D23} for pure $\rho_S$ and $\rho_E$, we obtain \eqref{D23} in this case.
Therefore, we obtain \eqref{5} in the case where $\rho_E$ is pure.

%%%%%%%%%%%%%%%%%

\bigskip

Finally, we show \eqref{5} for the case where $\rho_E$ is mixed.
We prove this in a similar manner to the proof of Theorem \ref{T1} in the case where $\rho_E$ is a mixed state.
We employ the decomposition of $\rho_E$ into pure states $\{ \rho_\eta\}$(i.e., $\rho_E=\sum_\eta q_\eta \rho_\eta$) as satisfying $V_{A_E}(\rho_E)=\sqrt{\sum_{\eta}q_\eta V_{A_E}(\rho_{\eta})^2}$.
We denote quantities, $\overline{\delta}(\rho_S, \{ \psi_i\} )$, $\delta (\rho_S)$ and $\delta (\psi_{i})$, with the initial state of $E$ as $\rho_{\eta}$ by putting the subscript $\eta$ such as $\overline{\delta}(\rho_S, \{ \psi_i\} )_{\eta}$, $\delta (\rho_S)_{\eta}$ and $\delta (\psi_{i})_{\eta}$.

Since \eqref{D23} has already been proven for a pure $\rho_E$, we have the following inequality holds for each $\rho_{\eta}$:
\begin{align}
V_{A_E}(\rho_\eta)\ge\frac{\chi(\rho_S, \{ \psi_i\} )}{5\sqrt{2}\times\overline{\delta}(\rho_S, \{ \psi_i\} )_{\eta}}-2\|A_S\|,\label{E28}
\end{align}
where $\overline{\delta}(\rho_S, \{ \psi_i\} )_{\eta}$ has the following expression
\begin{align}
\overline{\delta}(\rho_S, \{ \psi_i\} )_{\eta}=\sqrt{\delta (\psi_{i})^{2}_{\eta}+\sum_{i}r_i\delta (\rho_S)^{2}_{\eta}}.
\end{align}
with $r_i:=\sum_{\lambda}p_{\lambda}|\alpha^{(\lambda)}_{i}|^2$.
With keeping \eqref{E28} in mind, $V_{A_E}(\rho_E)$ is evaluated with a downward convex function $l(x):=(\max\{0,\frac{\chi(\rho_S, \{ \psi_i\} )}{5\sqrt{2}x}-2\|A_S\|\})^2$ ($x>0$) as
\begin{align}
\frac{1}{4}\calF(\rho_E)&=\sum_{\eta}q_{\eta}V_{A_E}(\rho_\eta)^2\nonumber\\
&\ge\sum_{\eta}q_{\eta}l(\overline{\delta}(\rho_S, \{ \psi_i\} )_{\eta})\nonumber\\
&\ge l(\sum_{\eta}q_{\eta}\overline{\delta}(\rho_S, \{ \psi_i\} )_{\eta}).\label{E30}
\end{align}

Here, we shall prove 
\eqa{
\sum_{\eta}q_{\eta}\overline{\delta}(\rho_S, \{ \psi_i\} )_{\eta}\le\sqrt{2}\times\overline{\delta}(\rho_S, \{ \psi_i\} ).
}{5-mix-mid3}
We start from the following simple inequality:
\begin{align}
\sum_{\eta}q_{\eta}\overline{\delta}(\rho_S, \{ \psi_i\} )_{\eta}&\le\sqrt{\sum_{\eta}q_{\eta}\overline{\delta}(\rho_S, \{ \psi_i\} )^{2}_{\eta}}\nonumber\\
&=\sqrt{\sum_{\eta}q_{\eta}\delta (\rho_S)^{2}_{\eta}+\sum_{\eta,i}q_{\eta}r_i\delta (\psi_{i})^{2}_{\eta}}. \lb{5-mix-mid2}
\end{align}
Here, the following inequality holds for any $\rho$:
\begin{align}
\sum_{\eta}q_{\eta}\delta (\rho)^{2}_{\eta}\le2\delta (\rho)^2, \lb{5-mix-mid}
\end{align}
which is proven as follows.
We denote the purification of $\rho$ by $\ket{\psi{SR}}$, and we define $\sigma_{SR}:=\Lambda_S(\psi_{SR})$ and $\sigma_{SR,\eta}$ as that with the initial state of $E$ as $\rho_\eta$.
We then have
\begin{align}
\left(1-\frac{(\delta (\rho))^2}{2}\right)^2
&=\bra{\psi_{SR}}U^{\dagger}_{S}\sigma_{SR}U_S\ket{\psi_{SR}}\nonumber\\
&=\sum_{\eta}q_{\eta}\bra{\psi_{SR}}U^{\dagger}_{S}\sigma_{SR,\eta}U_S\ket{\psi_{SR}}\nonumber\\
&=\sum_{\eta}q_{\eta}\left(1-\frac{\delta (\rho)^{2}_{\eta}}{2}\right)^2 \nt \\
&\le\left(1-\frac{\sum_{\eta}q_{\eta}\delta (\rho)^{2}_{\eta}}{2}\right)\nonumber\\
&\le\left(1-\frac{\sum_{\eta}q_{\eta}\delta (\rho)^{2}_{\eta}}{4}\right)^2,
\end{align}
which directly implies \eqref{5-mix-mid}.
Applying \eqref{5-mix-mid} to \eqref{5-mix-mid2}, we arrive at the relation \eqref{5-mix-mid3}.

Finally, substituting \eqref{5-mix-mid3} into \eqref{E30}, along with the non-increasingness of $l(x)$, we obtain
\begin{align}
\sqrt{\calF_{A_E}(\rho_E)}&\ge 2\sqrt{l(\sum_{\eta}q_{\eta}\overline{\delta}(\rho_S, \{ \psi_i\} )_{\eta})} \nonumber\\
&\geq 2\sqrt{l(\sqrt{2}\times\overline{\delta}(\rho_S, \{ \psi_i\} ))}\nonumber\\
&=\frac{\chi(\rho_S, \{ \psi_i\} )}{5\overline{\delta}(\rho_S)}-4\|A_S\| ,
\end{align}
which readily implies the desired result \eqref{5}.

\end{proofof}

\section{A generalized version of Theorem \ref{T1} for the case where $A_S+A_E$ is not conserved}
In this section, we consider the case that the total dynamics is unitary, but the quantity $A$ is not conserved perfectly, i.e., the case of $[U_{SE},A_S+A_E]\ne0$.
In this case, we can define the degree of asymmetry of $U_{SE}$:
\begin{align}
\calA_{U_{SE}}:=\frac{\lambda_{\max}([U_{SE},A_S+A_E])-\lambda_{\min}([U_{SE},A_S+A_E])}{2}.
\end{align}
Then, we can obtain the following theorem:
\begin{theorem}\label{T3}
When an implementation set $({\cal H}_{E}, A_E, \rho_E, U_{SE})$ implements $U_S$ within error $\delta $, the following inequality holds:
\begin{align}
\sqrt{\calF_{U_S,\delta}}\ge\frac{\calA_{U_S}-\calA_{U_{SE}}}{\delta}-6\max\{\|A_S\|,2\calA_{U_{SE}}\},\label{T3-2}
\end{align}
\end{theorem}

We can obtain the proof of this theorem just by substituting the following inequalities for \eqref{help1}, \eqref{help2} and $\Delta\le \|A_S\|$ in the proof of Theorem \ref{T1}:
\begin{align}
\max\{V_{A_E}(\sigma_{E,\uparrow}), V_{A_E}(\sigma_{E,\downarrow})\}&\le\delta+2\max\{\|A_S\|,2\calA_{U_{SE}}\},\label{help1b}\\
2(\calA_{U_S}-\calA_{U_{SE}})&\le\Delta+4\delta(\rho_{S,\uparrow+\downarrow})\|A_S\|.\label{help2b}\\
\Delta&\le \|A_S\|+\calA_{U_{SE}},\label{apparent}
\end{align}
We show these inequalities below.
The inequality \eqref{apparent} is obvious.
Let us show \eqref{help2b}.
We define
\begin{align}
\Delta^{S}_{\uparrow}&:=\Tr[A_S(\rho_{S,\uparrow}-\sigma_{S,\uparrow})],\\
\Delta^{S}_{\downarrow}&:=\Tr[A_S(\rho_{S,\downarrow}-\rho'_{S,\downarrow})],
\end{align}
Then, clearly $|\Delta^{S}_{\uparrow}-\Delta^{S}_{\downarrow}|\le \Delta+2\calA_{U_{SE}}$.
In the same manner as the derivation of \eqref{12.19.1}, we obtain
\begin{align}
2\calA_{U_S}\le|\Delta^{S}_{\uparrow}-\Delta^{S}_{\downarrow}|+4\delta(\rho_{S,\uparrow+\downarrow})\|A_S\|.
\end{align}
Therefore, we obtain \eqref{help2b}.

Next, we show \eqref{help1b}.
We use the following important fact:$\\$
\textit{
Let us take an arbitrary positive operator $A$ and arbitrary unitary $U$.
When $\|[U,A]\|\le\chi$ holds for a positive real number $\chi$, the following inequality holds for an arbitrary state $\rho$:}
\begin{align}
|V^{2}_{A}(\rho)-V^{2}_{U^{\dagger}AU}(\rho)|\le\chi(2V_{A}(\rho)+\chi),\label{fact}
\end{align}
\textit{where $V_{A}(\rho)$ is the standard deviation of $A$ in $\rho$.}$\\$
(\textit{Proof of \eqref{fact}:}
Because of $\|[A,U]\|=\|A-U^{\dagger}AU\|$, the Hermitian $X:=A-U^{\dagger}AU$ satisfies $\|X\|\le\chi$.
By using $X$, we can express $V^{2}_{U^{\dagger}AU}$ as follows:
\begin{align}
V^{2}_{U^{\dagger}AU}(\rho)&=\left<(A-X)^2\right>_{\rho}-\left<A-X\right>^{2}_{\rho}\nonumber\\
&=V^{2}_{A}(\rho)-2{\rm Cov}_{A;X}(\rho)+V^{2}_{X}(\rho),
\end{align}
where ${\rm Cov}_{A;X}(\rho):=\frac{1}{2}\Tr[\rho(AX+XA)]-\left<A\right>_{\rho}\left<X\right>_{\rho}$.
Because of $V_{X}(\rho)\le\|X\|\le\chi$ and the quantum correlation coefficient is lower than or equal to $1$, we obtain
\begin{align}
|V^{2}_{U^{\dagger}AU}(\rho)-V^{2}_{A}(\rho)|&\le2|{\rm Cov}_{A;X}(\rho)|+V^{2}_{X}(\rho)\nt\\
&\le2V_{X}(\rho)V_{A}(\rho)+V^{2}_{X}(\rho)\nt\\
&\le \chi(2V_{A}(\rho)+\chi).
\end{align}
\textit{(Proof end)})

Let us show \eqref{help1b}.
By using \eqref{fact}, we firstly show that the variances of $A_S+A_E$ in the initial and the final states are very close to each other.
The variance of $A_S+A_E$ in the initial state is $V^{2}_{A_S}(\rho_{S,i})+V^{2}_{A_E}(\rho_{E})$, and corresponds to $V^{2}_{A}(\rho)$ in \eqref{fact}.
The variance of $A_S+A_E$ in the final state is $V^{2}_{A_S}(\sigma_{S,i})+V^{2}_{A_E}(\sigma_{E,i})+2{\rm Cov}_{A_S+A_E}(e^{-iH\tau}(\rho_{S,i}\otimes\rho_{E})e^{iH\tau})$, and corresponds to $V^{2}_{U^{\dagger}AU}(\rho)$ in \eqref{fact}.
Substituting $A_S+A_E$, $e^{-iH\tau}$ and $\rho_{S,i}\otimes\rho_{E}$ for $A$, $U$ and $\rho$ of \eqref{fact}, we obtain 
\begin{align}
&V^{2}_{A_S}(\rho_{S,i})+V^{2}_{A_E}(\rho_{E})\nonumber\\
\ge&V^{2}_{A_S}(\sigma_{S,i})+V^{2}_{A_E}(\sigma_{E,i})\nonumber\\
&+2{\rm Cov}_{A_S+A_E}(e^{-iH\tau}(\rho_{S,i}\otimes\rho_{E})e^{iH\tau})\nonumber\\
&-\chi(2\sqrt{V^{2}_{A_S}(\rho_{S,i})+V^{2}_{A_E}(\rho_{E})}+\chi),
\end{align}
where $V_{A_S}(\rho)$ is the standard deviation of the quantity $A$ in $\rho$, and ${\rm Cov}_{A_S+A_E}(\sigma)$ is the covariance of $A$ of the state of $\sigma$ on $SE$.
Because $-V_{A_S}(\sigma_{S,i})V_{A_E}(\sigma_{E,i})\le {\rm Cov}_{A_S+A_E}(e^{-iH\tau}(\rho_{S,i}\otimes\rho_{E})e^{iH\tau})$ (this is a basic feature of the covariance) and $V_{S}(\rho)\le\|A_S\|/2$ for any $\rho$, we obtain
\begin{align}
&V_{A_E}(\sigma_{E,i})-V_{A_S}(\sigma_{S,i})\nonumber\\
&\leq \sqrt{V^{2}_{A_S}(\sigma_{S,i})+V^{2}_{A_E}(\sigma_{E,i})-2V_{A_S}(\sigma_{S,i})V_{A_E}(\sigma_{E,i})} \nonumber\\
&\leq \sqrt{V^2_{A_E}(\rho_{E})+V^2_{A_S}(\rho_{S,i})+\chi(2\sqrt{V^{2}_{A_S}(\rho_{S,i})+V^{2}_{A_E}(\rho_{E})}+\chi)}\nonumber\\
&\le V_{A_E}+1.5\max\{\|A_S\|,\chi\}.
\end{align}
By substituting $2\calA_{U_{SE}}$ for $\chi$, we obtain \eqref{help1b}.


\begin{thebibliography}{00}
\bibitem{footnote-1}Our results are valid even if the dynamics of $ SE $ is not unitary as long as the conservation law of energy is established. We discuss this extension in Appendix \ref{AppendixE}.
\bibitem{footnote0}When $[U_S,A_S]\ne0$, the inequality $\lambda_{\max}([U_S,A_S])>0>\lambda_{\min}([U_S,A_S])$ must hold, which follows from $\Tr[U^{\dagger}_SA_SU_S-A_S]=0$.
\bibitem{footnote} Let $S$ be a qubit, and set $A_S$ as $\ket{1}\bra{1}-\ket{0}\bra{0}$, $U_S$ as $\ket{1}\bra{0}+\ket{0}\bra{1}$, and $\rho_S$ as $(\ket{0}+\ket{1})/\sqrt{2}$.
We also set $E$ as a qubit, $A_E$ as $\ket{1}\bra{1}-\ket{0}\bra{0}$, $\rho_E$ as $\ket{0}\bra{0}$, and $U_{SE}$ as identity.
Then, both $V_{A'_S-A_S}(\rho_S)=1$, $\delta (\rho_S)=0$, $V_{A_E}(\rho_E)=0$ hold, which violates a relation in the anticipated form.
\bibitem{footnote3}Note that the magnitude of the quantum fluctuation in the external system $E$ is not necessarily greater than the expected value of $A_E$ itself.
Namely, the initial state $\rho_E$ might be satisfy 
$\Tr[\rho_EA_E]>>\calF_{A_E}(\rho_E)$.
\bibitem{footnote4}We can apply our results to the transient case that the final Hamiltonian of $S$ is different from the initial one with using the controller trick proposed in \cite{HO}.
Also, we can apply our results to the case that there is a third system which interacts the heat engine $S$ and work storage $E$. 
We discuss such cases in Appendix \ref{AppendixE}
\bibitem{footnote2} Here, we define $\calK\models_{\delta}\calE_S$ ($\calK=\calI$ or $\calJ$) as $F(\calE_S\otimes1_{R}(\psi_{SR}),\Lambda_S\otimes1_{R}(\psi_{SR}))\ge(1-\frac{\delta^2}{2})$ for arbitrary $\rho_S$. ($\psi_{SR}$ is a purification of $\rho_S$. This definition is independent of the choice of the purification $\psi_{SR}$.)
\bibitem{footnote-asymmetry}In general, the initial and final Hilbert space of ${\cal E}$ might be different.
In such a case, the equality \eqref{G-cov-def} becomes ${\cal E}(U_{g,\mathrm{in}}(...)U^{\dagger}_{g,\mathrm{in}})=U_{g,\mathrm{out}}{\cal E}(...)U^{\dagger}_{g,\mathrm{out}},\enskip \forall g\in G$.
Here $U_{g,\mathrm{in}}$ and $U_{g\mathrm{out}}$ are the unitary on the initial and final Hilbert spaces.
\end{thebibliography}
\end{document}